\documentclass[12pt]{article}
\usepackage{a4wide}
\usepackage{amssymb}
\usepackage{amsmath}
\usepackage{graphicx}
\usepackage{mathdots}
\usepackage{mathrsfs}
\usepackage{xcolor}
\usepackage{hyperref}

\def\makeatletter{\catcode`\@=11}
\makeatletter
\def\mathbox#1{\hbox{$\m@th#1$}}%
\def\math@ccstyles#1#2#3#4#5#6#7{{\leavevmode
      \setbox0\mathbox{#6#7}%
      \setbox2\mathbox{#4#5}%
      \dimen@ #3%
      \baselineskip\z@\lineskiplimit#1\lineskip\z@
      \vbox{\ialign{##\crcr
             \hfil \kern #2\box2 \hfil\crcr
             \noalign{\kern\dimen@}%
             \hfil\box0\hfil\crcr}}}}
\def\mathaccstyles{\math@ccstyles\maxdimen}
\def\maththroughstyles{\math@ccstyles{-\maxdimen}}
\def\unity%
 {\maththroughstyles{.45\ht0}\z@\displaystyle {\mathchar"006C}\displaystyle 1}

\usepackage{pifont}
\newcommand{\solidheart}{\ensuremath{\text{\ding{170}}}}

\begin{document}

\begin{flushright}\footnotesize

\texttt{}
\vspace{0.6cm}
\end{flushright}

\mbox{}
\vspace{0truecm}
\linespread{1.1}

\centerline{\LARGE \bf Boundary Layers and One-point Functions }
\medskip
\medskip
\centerline{\LARGE \bf in the Presence of Monodromy Defects  }
\medskip

\vspace{.5cm}

 \centerline{\LARGE \bf }

\vspace{1.5truecm}

\centerline{
    { \bf Hugo Calvo Castro${}^{a,\,b,\,\clubsuit}$},
    { \bf Ignacio Carre\~no Bolla${}^{a,\,b,\,\spadesuit}$},
    { \bf Diego Rodriguez-Gomez${}^{a,\,b,\,\solidheart}$}}

\begingroup
\renewcommand{\thefootnote}{}

\footnotetext[1]{$\clubsuit$ \texttt{calvohugo@uniovi.es}}
\footnotetext[2]{$\spadesuit$ \texttt{ignaciocarbolla@gmail.com}}
\footnotetext[3]{$\solidheart$ \texttt{d.rodriguez.gomez@uniovi.es}}

\endgroup

\vspace{1cm}
\centerline{{\it ${}^a$ Department of Physics, Universidad de Oviedo}} \centerline{{\it C/ Federico Garc\'ia Lorca  18, 33007  Oviedo, Spain}}
\medskip
\centerline{{\it ${}^b$  Instituto Universitario de Ciencias y Tecnolog\'ias Espaciales de Asturias (ICTEA)}}\centerline{{\it C/~de la Independencia 13, 33004 Oviedo, Spain.}}
\vspace{1cm}

\centerline{\bf ABSTRACT}
\medskip 

We study one-point functions of composites of charge $e$ operators in the presence of a monodromy defect for a $U(1)$ global symmetry with monodromy $\beta$. We first compute these in free massless and massive theories, recovering in the former case the known $\sin(e\pi\beta)$ dependence and obtaining in the latter a $\sin^2(e\pi\beta)$ dependence. We then turn to holography and compute 1-point functions for operators $O$ of charge $J=\Delta$ in $\mathfrak{su}(N)$ $\mathcal{N}=4$ SYM in the presence of a monodromy defect for a $U(1)\in SO(6)_R$. From a WKB analysis in large $\Delta$ we recover the structure of standard and anchored saddles previously found in the literature, finding that, to subleading order in $1/\Delta$, the anchored regime is resolved by a boundary layer effect. Finally, using heat kernel methods, we determine the monodromy dependence of the induced 1-point function for the composite $O^{\dagger}O$, finding a smooth $\sin^2(J\pi\beta)$ behavior.

\noindent 

\setcounter{footnote}{0}

\newpage

\tableofcontents

\section{Introduction and conclusions}

Besides the traditional local operators, Quantum Field Theories (QFT's) can contain extended objects supported on submanifolds of spacetime generically dubbed defects. These are very interesting as they exhibit further properties of the QFT otherwise invisible, such as central charges which typically satisfy monotonicity conditions. Moreover, defects play a central role in the modern description of symmetries. In this paper we will be interested on a particular class of defects dubbed monodromy defects. Monodromy defects are codimension 2 disorder operators which are defined through the monodromy that fields charged under a global symmetry undergo as they encircle the defect in the transverse space. Monodromy defects have attracted a lot of attention recently (see \text{e.g.} \cite{Rodriguez-Gomez:2026mjj,Bianchi:2021snj,Arav:2024exg,Bomans:2024vii,Conti:2025qwn,Conti:2025wyj,Conti:2025wwf,Copetti:2026ncv,Gomis:2025gzb} for a very partial list of recent references) as they play prominent roles in various contexts of interest.

In this paper we will consider monodromy defects associated to global 0-form $U(1)$ symmetries. Concentrating on planar defects inside (euclidean) QFT's defined on $\mathbb{R}^d$, we can split the space as

\begin{equation}
\label{coords}
ds^2=d\vec{x}^2=d\vec{\sigma}^2+dr^2+r^2d\theta^2\,,
\end{equation}
where $\vec{\sigma}$ labels the $\mathbb{R}^{d-2}$ worldvolume of the defect which is implicitly located at $r=0$. Then, local operators of charge $e$ under the $U(1)$ undergo a monodromy

\begin{equation}
    O(\theta+ 2\pi)= e^{i2\pi e\beta}\,O(\theta),\quad g \in G\,,
\end{equation}

Defects break the spacetime symmetries of the background, and thus allow to explore finer details of the theory. In particular, due to the breaking of rotational invariance $SO(d)\rightarrow SO(d-2)\times SO(2)$, otherwise forbidden 1-point functions are allowed. In the case of free theories, this has been discussed in \cite{Bianchi:2021snj}. In turn, this has been very recently explored from a holographic perspective in \cite{Rodriguez-Gomez:2026mjj}.\footnote{See also \cite{Linardopoulos:2026mut,Georgiou:2023yak}.} In that reference, the 2-point function of a particular class of operators charged under a $U(1)\in SO(6)_R$ of $\mathfrak{su}(N)$ $\mathcal{N}=4$ SYM is computed holographically in the presence of a monodromy defect for the $U(1)$. The operators considered in \cite{Rodriguez-Gomez:2026mjj} are giant gravitons.\footnote{See \cite{Bissi:2011dc,Holguin:2025dei,Anempodistov:2026dhi} for detailed discussion of correlation functions of giant gravitons.}  These belong to the family of 1/2 BPS operators in the $[0,J,0]$ of $SO(6)_R$ with charge $e=J$ under the $U(1)$ satisfying the BPS bound $\Delta=J$. In the case of $\Delta=J\sim N$ it is well-known that the operators are holographically best described as spherical D3 branes spinning in the background 10d geometry. Using this description, the 2-point function for giant gravitons (in a particularly symmetric kinematic regime) was computed in \cite{Rodriguez-Gomez:2026mjj}. From there, taking the coincidence limit and upon appropriate subtraction of the obvious contact divergence, the 1-point function for the operator $O^{\dagger}O(r)$ was computed. 

Since giant gravitons are operators of large dimension $\Delta\sim N$, they are best described as D3 branes, which wrap minimal surfaces in the holographic geometry. Upon integrating out the internal part of the geometry, the computation boils down to geodesics in the effective 5d SUGRA dual. In the absence of the monodromy defect, the 2-point function for the giant graviton is computed by a $U$-shaped geodesic --dubbed \textit{standard geodesic}. In turn, in the presence of the defect, an additional saddle appears corresponding to another $U$-shaped geodesic with the apex anchored at the $y_{\star}$ where the background ends  \cite{Rodriguez-Gomez:2026mjj} --hence the name \textit{anchored geodesic}. It is this anchored contribution what contains the 1-point of $O^{\dagger} O$. However, quite surprisingly, in the probe computation of \cite{Rodriguez-Gomez:2026mjj} the 1-point function discontinuously kicks in when the defect is included, resulting in a non-analyticity of the 1-point function in the monodromy parameter $\beta$. This is in sharp contrast with massless free field theory computation of \cite{Bianchi:2021snj}, where the 1-point function goes as $\sin(e\pi\beta)$ and thus smoothly vanishes as $\beta\rightarrow 0$. Interestingly, as we compute in this paper, in the massive case the 1-point function goes as $\sin^2(e\pi\beta)$. It was speculated in \cite{Rodriguez-Gomez:2026mjj} that this non-analytic behavior might be an artifact of the large $\Delta$ regime. The purpose of this paper is to study in closer detail such instance. To that matter we consider bulk mode dual to a generic $\Delta $ 1/2 BPS operator and compute the 1-point function (extracted from the suitably regularized coincidence limit of the 2-point function) focusing in particular in the limit of large $\Delta$. Quite surprisingly, while doing this in full detail is well beyond reach, since our ultimate goal is to capture the $\beta$-dependence of the 1-point function, it turns out that it is possible to extract the desired information. Our computation contains two relevant pieces of information. First, from a WKB analysis in large mass, the standard \textit{vs} anchored regimes naturally emerge. Moreover, as speculated in \cite{Rodriguez-Gomez:2026mjj}, a boundary layer effect forces the anchored regime not to actually reach $y_{\star}$ (see eq.\eqref{eq::ystarboundarylayer}). As speculated in \cite{Rodriguez-Gomez:2026mjj}, such boundary layer disappears in the strict large $\Delta$ limit. Second, we are able to compute the $\beta$-dependence of the 1-point function, finding that it is controlled by $\sin^2(e\pi\beta)$, qualitatively similar to the massive scalar case. Put together, these results support the picture advocated in \cite{Rodriguez-Gomez:2026mjj}. Namely, that a boundary layer effect, which becomes infinitesimally thin in the strict large $\Delta$ limit, smooths the non-analytic behavior of \cite{Rodriguez-Gomez:2026mjj} into $\sin^2(e\pi \beta)$.

An interesting question left open by our analysis is the $\sin^2(e\pi \beta)$ behavior of the 1-point function in the holographic and massive cases \textit{vs} the $\sin(e\pi \beta)$ of the free scalar case. A possible explanation for this difference is the special role of the defect operator $|\phi|^2$ in the free theory.\footnote{We thank Zohar Komargodski for suggesting this possibility to us.} As discussed in Appendix B of \cite{Barkeshli:2025cjs}, $|\phi|^2$ is exactly marginal at $\beta=0$, and the free theory admits two defect fixed points which collide at that point. As a consequence, the family of defect CFTs need not be analytic in $\beta$ near $\beta=0$. One way to see this is noticing that, due to the existence of the marginal operator $|\phi|^2$, the defect includes the insertion of $\int \lambda_{\star}|\phi|^2$, with $\lambda_{\star}$ determined, in conformal perturbation theory, from an equation of the form $0=\beta_{\lambda}\sim |\beta|\lambda+\lambda^2+\cdots$ (the first term coming from eq.B.4 of \cite{Barkeshli:2025cjs}), leading to $\lambda_{\star}\sim |\beta|$. Hence, varying the monodromy parameter along the family of defect fixed points inserts not only the integrated current, but also the marginal defect operator with coefficient $\partial_\beta\lambda_\star$. Since $\lambda_\star\sim |\beta|$ near the fixed-point collision, one has $\partial_\beta\lambda_\star\sim \operatorname{sgn}(\beta)$, so the variation of the defect is itself non-analytic at $\beta=0$. Thus, although charge conjugation still enforces invariance under $\beta\to-\beta$, the usual expectation that the leading correction should be quadratic no longer follows, providing a natural explanation for the $\sin(e\pi\beta)\sim |\beta|$ behavior of the free massless theory. In turn, away from the free point, $\phi^2$ ceases to be exactly marginal and the fixed-point collision is lifted. One then expects analyticity in $\beta$ to be restored, so that the leading correction is quadratic, yielding the $\sin^2(e\pi\beta)$ behavior observed in the massive and holographic cases.

 We have mostly concentrated on the dependence of the 1-point function on the monodromy parameter $\beta$. Yet,  the holographic background contains an additional degree of freedom $h$  that, upon proper quantization, on the dual CFT, corresponds to the choice of Levi subgroup of a Gukov-Witten defect \cite{Bomans:2024vii}. This degree of freedom is quite elusive in holography, since it is visible in the deep bulk of the SUGRA solution and becomes invisible in the  boundary. An interesting future direction would be to try to study the functional dependence of the 1-point function with a non-trivial quantized $h$ turned on using alternative techniques. Lastly, we remark that the most general Gukov-Witten defect involves turning on additional parameters other than $h$, but the price to pay is that the supergravity solution cannot be compactified to 5d and becomes a so-called bubbling geometry  \cite{Drukker:2008wr}, and has to be considered in the full 10d (see \cite{Choi:2024ktc,IzquierdoGarcia:2025jyb} for recent works).

In the rest of the paper we provide the details supporting the results claimed above. In section \eqref{sec::freefield} as a warm up we will study monodromy defects in free field theories. In the massless case we recover the result for the 1-point function in \cite{Bianchi:2021snj}, which goes as $\sin(e\pi \beta)$. In addition we also consider the case of a massive scalar, where the 1-point function goes as $\sin^2(e\pi\beta)$. In section \eqref{sec::holopraphy} we turn to the main subject of interest, namely monodromy defects for a $U(1)$ within the R-symmetry in $\mathfrak{su}(N)$ $\mathcal{N}=4$ SYM. Considering fluctuations dual to operators with charge $e=J$ under the $U(1)$ and dimension $\Delta=J$, a WKB analysis of the equation of motion naturally exhibits two regimes corresponding to the standard \textit{vs} anchored in \cite{Rodriguez-Gomez:2026mjj}, including the boundary layer. By using the heat kernel method we are able to compute the $\beta$ dependence of the 1-point function, finding the advertised $\sin^2(e\pi\beta)$ behavior. Finally, we compile in the appendix a number of details relevant for the computation.

\section{1-point functions in the presence of monodromy defects in free scalar theory} \label{sec::freefield}

As a warm up, let us consider a monodromy defect in a free field theory given by a scalar  $\phi$ charged under a $U(1)$ global symmetry with charge $q$ in Euclidean $\mathbb{R}^d$, with the following action

\begin{equation}
S=\int d^dx \; \left(|\partial \phi|^2+m^2|\phi|^2\right)\,.
\end{equation}
We add to this action a codimension 2 planar monodromy defect by imposing the boundary condition

\begin{equation}
    \phi(r,\theta+2 \pi)= e^{i 2 \pi e\beta } \phi(r,\theta)\,,
\end{equation}
with $(r,\theta)$ the coordinates of the transverse $\mathbb{R}^2$ and the defect is located at $r=0$. Alternatively, we could implement this condition by introducing a singular background gauge field for the $U(1)$ with the following profile 

\begin{equation}
    A= \beta\, d\theta\,,
\end{equation}
and minimally coupling it to the scalar. This gauge field can be gauged away with a singular gauge transformation at $r=0$, which implements the monodromy on the scalar.

We are interested in the Green's function $G(\vec{x}-\vec{x}')=\langle \phi^{\dagger}(\vec{x})\phi(\vec{x}')\rangle$ in the presence of the defect. As reviewed in Appendix \eqref{sec::2-pointfunction}, one finds

\begin{align}
&G=\int \frac{d^{d-2}\vec{p}}{(2\pi)^{d-2}} \sum_{n\in \mathbb{Z}}\,\frac{1}{2\pi}\,e^{-in(\theta-\theta')}\,e^{i\vec{p}\cdot(\vec{\sigma}-\vec{\sigma}')}\,G_{n,\vec{p}}(r,r')\,,\\ \nonumber \\
&G_{n,\vec{p}}(r,r')=-I_{|n-e\beta|}(\omega r_<)\,K_{|n-e\beta|}(\omega r_>)\,,\qquad r_<={\rm min}(r,r')\,,\qquad r_>={\rm max}(r,r')\,.
\end{align}

\subsection{One-point function}
 
The presence of a defect breaks Lorentz symmetry and allows local operators to take VEV. We can read-off this VEV from the coincidence limit $\vec{x}'\rightarrow \vec{x}$ of the 2-point function upon removing the obvious contact divergence. To do so, we can simply subtract the Green's function without the defect (i.e. by setting $\beta=0$). This procedure can be thought of as the typical subtraction of the bulk degrees of freedom to study the genuine defect degrees of freedom, as usual in defect literature. The one point function is then

 \begin{equation}
 \begin{aligned}
     \langle \phi^\dagger \phi(r)\rangle&= \lim _{r'\to r}\langle \phi^\dagger(r) \phi(r')\rangle-\langle \phi^\dagger(r) \phi(r')\rangle_{\beta=0}=\\
     &=\frac{1}{2 \pi}\int \frac{d^{d-2}\vec{p}}{(2\pi)^{d-2}}\, \sum_{n\in \mathbb{Z}}I_{|n-e\beta|}(\omega r)\,K_{|n-e\beta|}(\omega r)-I_{|n|}(\omega r)\,K_{|n|}(\omega r)\,.
 \end{aligned}
 \end{equation}

\subsubsection{Zero mass limit}

As a check, we would like to recover the CFT limit of \cite{Bianchi:2021snj} by setting $m=0$. The 1-point function in that limit is (the details of the computation are shown in Appendix \eqref{sec::one-pointfunction})

\begin{equation}
\langle \phi^\dagger \phi(r)\rangle=\frac{1}{2(2\pi)^{\frac{d}{2}}r^{d-2}}\, \int_0^{\infty} dx\,x^{\frac{d}{2}-2}\,e^{-x} \sum_{n\in \mathbb{Z}}I_{|n-e\beta|}(x)-I_{|n|}(x)\,.
\end{equation}
The sum can be written as

\begin{equation}\label{eq::fullsumI}
\sum_{n\in \mathbb{Z}}I_{|n-e\beta|}(x)-I_{|n|}(x)=\tilde{C}  - \frac{\sin(e\pi \beta)}{\pi}\int_0^{\infty}dt\,e^{-x\cosh t} \frac{e^{e\beta t}+e^{(1-e\beta)t}}{1+e^t} \,,
\end{equation}
where 

\begin{equation}
\tilde{C}=\frac{1}{2\pi}\int_{-\pi}^{\pi}d\phi\,e^{x\cos\phi} \sum_n \left( e^{i|n-e\beta|\phi}-e^{i|n|\phi}\right)\,.
\end{equation}

Let us drop for now $\tilde{C}$. Inserting the result for the sum into the expression for the 1-point function we find

\begin{equation}
   \langle \phi^\dagger \phi(r)\rangle= -\frac{\sin(e\pi \beta)}{(2\pi)^{\frac{d}{2}+}r^{d-2}}\, \int_0^{\infty}dt\, \frac{e^{e\beta t}+e^{(1-e\beta)t}}{1+e^t}\int_0^{\infty}\, dx\,x^{\frac{d}{2}-2} \,e^{-x(1+\cosh t)}  \,.
\end{equation}
All integrals can now be done. When the dust settles we find

 \begin{equation} \label{eq::massless1pt}
     \langle \phi^\dagger \phi(r)\rangle=-\frac{ \sin
   (e\pi  \beta) \Gamma \left(\frac{d}{2}-e\beta\right) \Gamma
   \left(\frac{d}{2}+e\beta-1\right)}{2^{d-1}  \pi ^{\frac{d+1}{2}}(d-2) \Gamma
   \left(\frac{d-1}{2}\right)} \frac{1}{r^{d-2}}\,,
 \end{equation}
 which reproduces the result of \cite{Bianchi:2021snj}.

 Let us now come back to the dropped $\tilde{C}$ term. Inserting it into the expression for the 1-point function, if would give a contribution
 
 \begin{equation}
     C=\frac{\pi}{(2\pi)^{\frac{d}{2}}r^{d-2}}\int_{-\pi}^{\pi}d\phi\,\sum_n \left( e^{i|n-e\beta|\phi}-e^{i|n|\phi}\right)\int_0^{+\infty} \,dx\,x^{\frac{d}2-2}e^{-x(1-\cos\phi)} \,.
 \end{equation}
Doing the $x$-integral we find
  
  \begin{equation}
     C=\frac{\pi \,\Gamma(\frac{d}{2}-1)}{(2\pi)^{\frac{d}{2}}r^{d-2}}\int_{-\pi}^{\pi}d\phi\,\frac{\sum_n \left( e^{i|n-e\beta|\phi}-e^{i|n|\phi}\right)}{(1-\cos\phi)^{\frac{d}{2}-1}}\,.
 \end{equation}
This integrand has a pole at $\phi=0$, and so it will be divergent. With the appropriate renormalization prescription, we can indeed neglect it.

\subsubsection{Large mass limit}
 
Let us now consider the $m\to \infty$ limit. In this case, the 1-point function becomes (the details of the computation are shown in Appendix \eqref{sec::one-pointfunction})

\begin{equation}
\begin{aligned}
    \langle \phi^\dagger \phi(r)\rangle&=\frac{ m^{\frac{d}{2}-1}}{(4 \pi)^{\frac{d}{2}}}\sum_{n\in\mathbb{Z}}\,\frac{\,e^{-mS_0}}{z_*^{\frac{d}{2}} (\det H)^\frac12}\,(e^{i2\pi ne \beta}-1)=\\=& 
  \sqrt{\frac{\pi}{2}} \frac{m^{\frac{d}{2}-1}}{(2 \pi)^d \,r^{\frac{d-1}{2}}}\sum_{n\in\mathbb{Z}}|n|^{\frac{1-d}{2}}
   e^{-2 \pi  m |n| r}(e^{i2\pi ne \beta}-1)\,.
\end{aligned}
\end{equation}
Note that since $m$ is large, because of the factor $e^{-2 \pi  m |n| r}$ the main contribution comes from $n=\pm 1$.

\begin{equation}
    \langle \phi^\dagger \phi(r)\rangle=  \sqrt{\frac{\pi}{2}} \frac{m^{\frac{d}{2}-1}}{(2 \pi)^d } \frac{e^{-2\pi m r}\,\sin^2(e\pi \beta)}{r^{\frac{d-1}{2}}\,}\,.
\end{equation}

\section{1-point functions in the presence of monodromy defects in $\mathcal{N}=4$ SYM from holography} \label{sec::holopraphy}

After our free field theory warm-up, let us now turn to the object of interest, namely the 1-point functions of giant gravitons activated by the defect in $\mathcal{N}=4$ SYM from holography, which can be read-off from the coincidence limit of the 2-point function upon appropriate regularization. The first step is to introduce the holographic backgrounds of interest. These arise as a solution to 5d gauged supergravity, proposed originally in \cite{Kunduri:2007qy,Ferrero:2021etw} and recently studied in \textit{e.g.} \cite{Arav:2024exg,Bomans:2024vii,Conti:2025qwn,Conti:2025wyj,Conti:2025wwf} . The background has three $U(1)$ gauge fields coming from truncating to the Cartans of the full R-symmetry $ SO(6)$. We will only turn on a diagonal $U(1)$ for simplicity. The explicit form of the background is 

\begin{align}
& ds^2=H^{\frac{1}{3}}\,\Big[ds_{AdS_3}^2+\frac{1}{4P}\,dy^2+\frac{P}{H}\,d\theta^2\Big]\,,\qquad A=(\alpha-\frac{y}{y+q})\,d\theta\,,\nonumber \\ 
& P=H-y^2\,,\qquad H=(y+q)^3\,,\nonumber \\
& q=(h+\beta)(1-h-\beta)^2,\quad \alpha=1-\beta\,.
\end{align}

The holographic boundary is at $y=+\infty$. Upon doing the change \cite{Rodriguez-Gomez:2026mjj}

\begin{equation}
(u,\,y,\, \tan\theta)\,\rightarrow \, (\sqrt{r^2+z^2},\, \frac{\vec{x}^2+z^2}{z^2},\frac{x_2}{x_1})\,,
\end{equation}
the geometry becomes $AdS_5$. In particular, note that close to the holographic boundary $y\sim z^{-2}\,\vec{x}^2$, being $\vec{x}^2$ the position in the $\mathbb{R}^2$ transverse to the defect. Moreover, asymptotically  $A=-\beta d\theta$, which shows that $\beta$ is precisely the monodromy parameter associated to the defect. In turn, the $y$ coordinate is defined up to a minimal $y_{\star}$ defined as

\begin{equation}
    y_*=(1-(\beta+h))^3\,.
\end{equation}
Expanding $y=y_*+\rho^2$, the metric becomes

\begin{align}
    & ds^2\sim\,(1-(h+\beta))^2\,\Big( ds_{AdS_3}^2+R\,\big(d\rho^2+T^{-2} \rho^2\,d\theta^2\big)\Big)\,,\nonumber\\
    &R=\left|\frac{1}{(1-(h+\beta))^3\,(1-3(h+\beta))}\right|\,,\qquad  T=\left|\frac{1}{1-3(\beta+h)}\right|\,.
\end{align}

The particular defect described by the background depends on the choice of the parameters $\beta$ and $h$. For instance, as discussed in \cite{Rodriguez-Gomez:2026mjj}, setting $h=0$ corresponds to the simplest case where the defect is only labelled by its monodromy $\beta$ and carries no other degree of freedom. On the other hand \cite{Bomans:2024vii}, turning on both $\beta$ and $h$ and quantizing $T=k \in \mathbb{Z}$ corresponds to a choice of Levi subgroup for a specific type of Gukov-Witten defect. \footnote{Gukov-Witten defects in general allow for more generic choices of Levi subgroup than what we have here. In order to allow this more general case, we need to consider the bubbling geometries of \cite{Drukker:2008wr}, very recently re-considered in \cite{Choi:2024ktc,IzquierdoGarcia:2025jyb}.}

\subsection{Holographic correlators in the presence of the defect and the 1-point function}

We are interested in computing holographically the 2-point function for giant gravitons in the presence of the monodromy defect, specially in the coincidence limit where we can read-off the 1-point function of the giant graviton activated by the defect. To that matter, we should first identify the supergravity fluctuation dual to the giant graviton. The giant graviton belongs to the 1/2 BPS family in the $[0,J,0]$ of $SO(6)_R$ with $\Delta=J$ in the particular regime $J\sim N$. It is natural to guess that the supergravity fluctuation dual to a generic member of the family is a complex scalar of mass $m$ (related in the usual way to $\Delta$) coupled to the R-symmetry gauge field with charge $e=J$. The action is generically

\begin{equation}
S=\int d^5x\, \sqrt{g}\,(|D\Phi|^2+m^2\,|\Phi|^2)\,,\qquad D_M\phi=\partial_M\Phi-i\,e\,A_M\,\Phi\,.
\end{equation}
Since we are interested in describing giant gravitons, we will be interested in the limit of large $\Delta$, when $m\sim \Delta=J=e\gg 1$. In particular, we will use interchangeably $m$ and $\Delta$.

As reviewed in Appendix \eqref{sec::2pointreview}, the 2-point function can be read-off in the standard way from the solutions to the equation of motion. Then, just as in the field theory exercise, we could read-off the activated VEV's by considering the appropriately regulated coincidence limit. Thus, our immediate task is to solve the bulk equation of motion. Using the explicit form of the background, it reads

\begin{equation}
H^{-\frac{1}{3}}\,\Box_{AdS_3}\Phi+4\,H^{-\frac{1}{3}}\,\partial_y\big(P\,\partial_y\Phi\big)+\frac{H^{\frac{2}{3}}}{P}\,\big(\partial_{\theta}-i eA\big)^2\Phi-m^2\Phi=0\,,
\end{equation}
where

\begin{equation}
\Box_{AdS_3}\Phi=u^3\,\partial_u\left(\frac{\partial_u\Phi}{u}\right)+u^2\,(-\partial_t^2\Phi+\partial^2_x\Phi)\,.
\end{equation}

We expand

\begin{equation}
\Phi=\int d\lambda\,\mu(\lambda)\,\sum_{n\in \mathbb{Z}} e^{-in \theta}\,u_{\lambda}\,\phi_{\lambda,n}(y)\,,
\end{equation}
where $u_{\lambda}$ are the $AdS_3$ eigenfunctions satisfying

\begin{equation}
\Box_{AdS_3}u_{\lambda}=\lambda\,(\lambda-2)u_{\lambda}\,;
\end{equation}
and  $\mu(\lambda)$ is a measure for the $AdS_3$ modes (which will be largely irrelevant for us). Then, the equation of motion boils down to

\begin{equation}
\label{eq::radial}
4\,H^{-\frac{1}{3}}\,\partial_y\big(P\,\partial_y\phi_{\lambda,n}\big)-\left[H^{-\frac{1}{3}}\,\lambda\,(\lambda-2)+m^2+\frac{H^{\frac{2}{3}}}{P}\,(n+e\,A)^2\right]\phi_{\lambda,n}=0\,.
\end{equation}

\subsection{The WKB approximation}

Solving the mode equation above is very hard. Fortunately, we are interested in the regime of large $m=e\sim \Delta\sim N$. Owing to this, we can attempt to solve this equation \textit{via} the WKB approximation on the mass. It is useful to first do the field redefinition $\phi_{\lambda,n-e\beta}=P^{-\frac12} \psi_{\lambda,n-e\beta}$ into the Schrödinger-like equation

\begin{equation}
\partial_y^2\psi_{\lambda,n}-V(y)\psi_{\lambda,n-e\beta}=0\,,\qquad V(y)=\frac{1}{4\,P} \Big[\lambda\,(\lambda-2)+H^{\frac{1}{3}}m^2-\frac{P'^2-2\,P\,P''}{P}+\frac{H}{P}\,(n+e\,A)^2\Big]\,,
\end{equation}
where $m^{-1}$ plays the role of $\hbar$.  We then write the semiclassical wavefunction $\psi_{\lambda,n}=e^{-m S_{\lambda,n}}$, so that to leading order in $m$

\begin{equation}
(S_{\lambda,n}')^2= V\,.
\end{equation}
For giant gravitons $m\sim e$. Moreover, it is useful to redefine parameters as follows

\begin{equation}
  \lambda= \xi m,\quad n= r m\,.
\end{equation}
Then, the potential $V$ becomes

\begin{equation} \label{eq::WKBpotential}
   V(y)=\frac{1}{4\,P^2} \Big[(H^{\frac{1}{3}}+\xi^2)\,P+H\,(r+A)^2\Big]-\frac{P'^2-2\,P\,P''}{4\,P^2 \,m^2}\,.
\end{equation}
The second term is subleading in $m$, and thus it can be dropped. As usual in WKB, the solution is obtained by matching asymptotic regions. In the large $y$ region we find
 
\begin{equation}
    V \sim \frac{1}{4 y^2},\quad \phi_{n,\lambda}^{UV}=A_{UV}\,y^{-\frac{m+3}{2}}+B_{UV}\,y^{\frac{m-3}{2}}\,.
\end{equation}
In turn, for small $y$, as standard in WKB, the solution will have a turning point $y_{\rm m}$ when $V(y_{\rm m})=0$. Explicitly, the location of the turning point is

\begin{equation} \label{eq::tphologr}
    (H^{\frac{1}{3}}+\xi^2)\,P+c^2\,H\,(r+A)^2=0\,.
\end{equation}
Two cases, mimicking the geodesic analysis of \cite{Rodriguez-Gomez:2026mjj}, now appear

\begin{enumerate}
    \item \textbf{Standard regime:} for generic values of $r,\,\xi$, the solution to \eqref{eq::tphologr} is some regular point in the bulk $y_{\rm m}>y_*$. Near the turning point the potential admits a Taylor expansion of the form
    
    \begin{equation}
V=V_1(r,\xi)\,(y-y_{\rm m})+\cdots\,,
\end{equation}
with $V_1(r,\xi)$ some function which includes the $n-e\beta$ dependence and whose exact form will not matter. The WKB equation near this point is solved by

\begin{equation}
(S_{\lambda,n-e\beta}')^2=V_1(r,\xi)\,(y-y_{\rm m})\qquad\leadsto \qquad S_{\lambda,n-e\beta}=c_{\rm n-e\beta}\pm \frac{2}{3}\,\sqrt{V_1}\,(y-y_{\rm m})^{\frac{3}{2}}\,.
\end{equation}
It is worth emphasizing that $P(y_{\rm m})\neq 0$, and consequently the neglected $\mathcal{O}(1/m^2)$ term in \eqref{eq::WKBpotential} can indeed be safely neglected.

\item \textbf{Anchored regime:} if $r=\beta-h$, the second term of \eqref{eq::tphologr} vanishes at $y=y_*$ and the turning point is exactly at $y_{\rm m}=y_*$, where the potential $V$  diverges. Not only that, but also $P(y_{\star})=0$, and consequently the $\mathcal{O}(1/m^2)$ term in \eqref{eq::WKBpotential}  cannot be neglected. In fact, keeping this term we see that the turning point acquires a $\mathcal{O}(m^{-2})$ correction

\begin{equation}
\label{eq::ystarboundarylayer}
    y_{\rm m}=(1-(\beta+h))^3+\frac{1}{m^2} \frac{(1-(\beta+h))^3 (1-3(\beta+h))}{(1-(\beta+h))^2+\xi ^2}\,+\cdots\,.
\end{equation}
As a consequence, the anchored regime never actually reaches $y_*$ and instead stops at an $O(m^{-2})$ distance. As this corrected $y_{\rm m}$ is a regular point in the bulk, we expect the solution to show an analytic behavior. This in turn strongly suggests that the non-analyticity present in \cite{Rodriguez-Gomez:2026mjj} is just an artifact of the approximation of taking first the strict large $m$ limit, effectively erasing this boundary layer effect. 

\subsection{The heat kernel}

In the previous section we have seen that the WKB approach allows us to recover the qualitative picture advocated in \cite{Rodriguez-Gomez:2026mjj}, including the anchored regime and its resolution through a boundary layer effect. The apparent non-analyticity in $\beta$ of the 1-point function in \cite{Rodriguez-Gomez:2026mjj} appears then as an artifact of the strict large $m$ limit assumed by the geodesic approximation, where the boundary layer effect is infinitely compressed. However, explicitly finding the VEV --or more precisely, its analytic $\beta$ dependence-- seems extremely hard. Yet, as reviewed in Appendix \eqref{sec::2pointreview}, we can find a direct path towards the 2-point function starting with \eqref{eq::osactiontouse}. Since $z\sim r\, y^{-\frac{1}{2}}$

\begin{equation}
\langle O^{\dagger}(\vec{x}')O(\vec{x})\rangle=\mathcal{N}^2\,\mathcal{G}(\vec{x}',\vec{x})\,,\qquad \mathcal{G}(\vec{x}',\vec{x})=r^{-\Delta}\,r'^{-\Delta}\, \lim_{y,y'\rightarrow \infty}\,y^{\frac{\Delta}{2}}\,y'^{\frac{\Delta}{2}}\,G(y',\vec{x}';y,\vec{x})\,,
\end{equation}
where $G$ is the Green's function for the bulk equation of motion. Expanding $G$ in modes

\begin{equation}
\langle O^{\dagger}(\vec{x}')O(\vec{x})\rangle=|\vec{x}|^{-\Delta}\,|\vec{x}'|^{-\Delta}\,\mathcal{N}^2\,\lim_{y,y'\rightarrow \infty}\,\,\int d\lambda\, \mu(\lambda) \, \left(y^{\frac{\Delta}{2}}\,y'^{\frac{\Delta}{2}}\,u_{\lambda}\right)\,\sum_{n\in \mathbb{Z}}\,e^{in(\theta-\theta')}\, \mathcal{G}_{\lambda,n-eA}\,,
\end{equation}
where $\mathcal{G}_{\lambda,n-eA}$ stands for the propagator of the purely radial equation \eqref{eq::radial}.\footnote{We include in the notation the $A$ label to remind the reader that such propagator is to be computed with the $A$ corresponding to the defect.} Just as before, evaluating exactly these expressions looks like a daunting task. However, we are mostly interested in extracting the VEV --more precisely its $\beta$ dependence-- activated by the presence of the defect. Such dependence is clearly solely contained in the angular sum of the radial Green's function, for which it is enough to compute the bulk radial Green's function. In turn, this propagator can be written in terms of the heat kernel in the standard manner as (we briefly review the heat kernel method in Appendix \eqref{sec::heatkern})

\begin{equation}
\mathcal{G}_{\lambda,n-eA}(X,Y)=\int_0^{\infty} ds\,e^{-\epsilon s}\,\mathcal{K}_{\lambda,n-eA}(s,X,Y)\,,\qquad (\partial_s+\Box_{\rm bulk})\,\mathcal{K}_{\lambda,n-eA}=0\,,
\end{equation}
where $\Box_{\rm bulk}$ is the radial bulk equation of motion \eqref{eq::radial}, which depends on the $\lambda$ and $n-eA$. For reasons that will become clear shortly, we have also introduced a short $s$ regulator $\epsilon$.

Moreover, since we are mostly interested in the activated VEV --more precisely, on its $\beta$ dependence--, we must take the the coincidence limit in the 2-point function. As it is well-known, the coincidence limit of the heat kernel corresponds to the short $s$ region, where it typically simplifies, giving us a hope of success. Encouraged by these observations, let us look for an expression of for the activated 1-point functions --more precisely, of their $\beta$ dependence-- in the presence of the defect. Just as in our free field theory warm-up, we must to remove contact divergences by subtracting the coincidence limit of the giant graviton 2-point function without the defect. Hence 

\begin{equation}
\langle O O^\dagger (r) \rangle=\frac{1}{r^{2\Delta}}\,\times\,\int d\lambda\,\mu(\lambda)\,\sum_{n\in\mathbb{Z}} \left( e^{-i n \epsilon}\,\mathcal{G}_{\lambda,n-eA}(\vec{x}+\epsilon;\vec{x})-e^{-in \epsilon }\,\mathcal{G}_{\lambda,n}(\vec{x}+\epsilon;\vec{x})\right)\,,
\end{equation}
where we are regulating the coincidence limit as $x'_i=x_i'+\epsilon$, and we have denoted $\mathcal{G}_{\lambda,n-eA}$ the 2-point function to indicate that it is to be computed in the presence of the corresponding gauge field ($A$ in the presence of the monodromy, 0 in the absence subtracted). Moreover, we are reabsorbing in $\mu(\lambda)$ prefactors which are irrelevant to obtain the $\beta$ dependence of the VEV. Reassuringly, the expected position-dependence fixed by conformal invariance emerges.

In the coincidence limit indeed the heat kernel dramatically simplifies. As reviewed in Appendix \ref{sec::heatkern}, assuming the adiabatic approximation that the background functions vary slowly, we can write

\begin{equation}
\mathcal{K}_{\lambda,n-eA}=\frac{1}{\sqrt{s}}\,e^{-s\,\frac{H}{P}\,(n-eA)^2}\,K_{\lambda}\,,\qquad K_{\lambda}=e^{-s\,(\lambda\,(\lambda-2)+m^2)}\,,
\end{equation}
Plugging this back into the Green's function,

\begin{align}
\label{eq::intermediate}
 \langle O O^\dagger (r) \rangle&=\frac{1}{r^{2\Delta}}\,\times\,\int_0^{+\infty} \frac{ds}{\sqrt{s}}\, \int d\lambda\,\mu(\lambda)\,K_{\lambda}\,e^{-\epsilon s}\,\sum_{n\in\mathbb{Z}} \left( e^{-i n \epsilon}\,e^{-s\,\frac{H}{P}\,(n+eA)^2}-e^{-in \epsilon }\,e^{-s\,\frac{H}{P}\,n^2}\right)\nonumber \\ & \sim \frac{1}{|\vec{x}|^{2\Delta}}\,\times\,
\int_0^{+\infty} \frac{ds}{\sqrt{s}}\, \int d\lambda\,\mu(\lambda)\,K_{\lambda}\,e^{-\epsilon s}\,\sum_{n\in\mathbb{Z}} \left( e^{-s\,\frac{H}{P}\,(n+eA)^2}-\,e^{-s\,\frac{H}{P}\,n^2}\right)\,.
\end{align}

Doing now Poisson-Jacobi summation we find

\begin{equation}
\langle O O^\dagger (r) \rangle= \frac{1}{r^{2\Delta}}\,\times\, \int d\lambda\,\mu(\lambda)\,\sum_{n\in\mathbb{Z}} \Big\{\int_0^{+\infty} \frac{ds}{s}\,\sqrt{\frac{P}{2 H}}\,e^{-s(\lambda(\lambda-2)+m^2)}\,e^{-\frac{P}{s\,H}\pi^2n^2}\Big\}\, \left(e^{-i2\pi e\,A\,n}-1\right)\,.
\end{equation}

We now need to do the integral in brackets. To do so, it is useful to redefine $\lambda=m\hat{\lambda}$, $n=m\hat{n}$, $\tau=m\,s$, so that

\begin{equation}
\Big\{ \cdots \Big\}= \int _0^{+\infty}\,d\tau f(\tau) e^{-m S(\tau)}\,, \qquad f(\tau)=\sqrt{\frac{P}{2\,H}}\frac{1}{\tau},\quad S(\tau)= \tau(\hat{\lambda}^2+1)+\frac{P}{\tau\,H}\pi^2\hat{n}^2\,.
\end{equation}
Since $m$ is large we can do the integral using the saddle-point approximation. The saddle point is

\begin{equation}
    \tau_*=\frac{\pi  |n| \sqrt{P}}{\sqrt{H
   \left(\hat{\lambda}^2+1\right)}}, \quad e^{-m S_*}=e^{-\frac{2 \pi  m |n| \sqrt{P} \sqrt{
   \hat{\lambda}^2+1}}{\sqrt{H}}}\,.
\end{equation}
Plugging the saddle point value of the integral, the 1-point function becomes

\begin{equation}
\langle O O^\dagger (r) \rangle\sim \frac{1}{r^{2\Delta}}\,\times\, \int d\lambda\,\mu(\lambda)\,\sum_{n\in\mathbb{Z}} \,\frac{ e^{-\frac{2 \pi  m |n| \sqrt{P} \sqrt{
   \hat{\lambda}^2+1}}{\sqrt{H}}}}{\sqrt{|n|}}\left(e^{-i2\pi eA\,n}-1\right)\,,
\end{equation}
where we are dropping numerical prefactors. Note that this is to be evaluated at large $y$, where $A=-\beta$. It yet remains to do the sum and the $\lambda$ integral. However, due to the overall exponential suppression, the sum is dominated by the $n=\pm 1$ terms. Thus, we can approximate the result, up to factors, as

\begin{equation}
    \langle O O^\dagger (r) \rangle \sim\frac{\sin^2{(e \pi \beta)}}{r^{2\Delta}}\,.
\end{equation}

Note that up to this point we have not needed to use that $e-J\sim \Delta$ is large. Thus, one might be afraid of directly extending our result to $e=J$. To analyze the case where $e=J\sim \Delta$, let us go back to \eqref{eq::intermediate}, which we reproduce for convenience 

\begin{equation}
 \langle O O^\dagger (r) \rangle= \frac{1}{|\vec{x}|^{2\Delta}}\,\times\,
\int_0^{+\infty} \frac{ds}{\sqrt{s}}\, \int d\lambda\,\mu(\lambda)\,e^{-s\,(\lambda\,(\lambda-2)+m^2)}\,e^{-\epsilon s}\,\sum_{n\in\mathbb{Z}} \left( e^{-s\,\frac{H}{P}\,(n+eA)^2}-\,e^{-s\,\frac{H}{P}\,n^2}\right)\,.
\end{equation}
In large $m$, and at large $y$ it can be massaged into

\begin{align}
 \langle O O^\dagger (r) \rangle & = \frac{1}{|\vec{x}|^{2\Delta}}\,\times \nonumber \\ &
 \int d\lambda\,\mu(\lambda) \sum_{n\in\mathbb{Z}} \left( \int_0^{+\infty} \frac{ds}{\sqrt{s}}\, e^{-\epsilon s} e^{-s\,m^2\,\left((\frac{n}{m}+\beta)^2+\frac{\lambda^2}{m^2}+1\right)}-\int_0^{+\infty} \frac{ds}{\sqrt{s}}\, e^{-\epsilon s} e^{-s\,m^2\,\left(\frac{n^2}{m^2}+\frac{\lambda^2}{m^2}+1\right)}\right)\,.
\end{align}
The integrals can be easily done

\begin{align}
 \langle O O^\dagger (r) \rangle  = \frac{1}{|\vec{x}|^{2\Delta}}\, \int d\lambda\,\mu(\lambda) \frac{\sqrt{\pi}}{m}\,\sum_{n\in\mathbb{Z}} f(\frac{n}{m}-\beta)-f(\frac{n}{m})\,,\qquad f(x)=\frac{1}{\sqrt{ x^2+\frac{\lambda^2}{m^2}+1}} \,.
\end{align}
Using that

\begin{equation}
\int dx\, \frac{e^{-i2\pi n x}}{\sqrt{ (\frac{x}{m}+\beta)^2+\frac{\lambda^2}{m^2}+1}}=2m\,e^{i2\pi m n\beta}\, K_0\left(2\pi |n| m \sqrt{\frac{\lambda^2}{m^2}+1}\right)\,,
\end{equation}
we can do Poisson resummation of the sum in $n$

\begin{align}
 \langle O O^\dagger (r) \rangle  = \frac{2\sqrt{\pi}}{|\vec{x}|^{2\Delta}}\, \int d\lambda\,\mu(\lambda) \sum_{k\in\mathbb{Z}} (e^{i2\pi m k \beta}-1)\, K_0\left(2\pi |k| m \sqrt{\frac{\lambda^2}{m^2}+1}\right) \,.
\end{align}
For large $m$ the Bessel function can be approximated as

\begin{equation}
K_0\left(2\pi |k| m \sqrt{\frac{\lambda^2}{m^2}+1}\right)\sim \frac{1}{\sqrt{4|k|m\sqrt{\frac{\lambda^2}{m^2}+1}}}\,e^{-2\pi |k| m\sqrt{\frac{\lambda^2}{m^2}+1}}\,.
\end{equation}
Hence, due to the exponential suppression, once again only the $k=\pm 1, 0$ terms contribute, from which the $\sin^2 (J\pi \beta)$ behavior is reproduced.

\end{enumerate}

\section*{Acknowledgments}

We would like to thank Zohar Komargodski for very useful conversations. I.C.B. would like to thank Andrea Conti and Ricardo Stuardo for useful discussions. I.C.B. would like to thank IPhT Saclay for the kind hospitality while part of this work was being carried out. I.C.B. would also like to thank the COST
Action CA22113 ``Fundamental challenges in theoretical physics” for financial support during part of this work through an STSM Grant. H.C.C., I.C.B. and D.R.-G are supported in part by the Spanish national grant MCIU-22-PID2021-123021NB-I00.  I.C.B. is also supported by the Severo Ochoa fellowship NAC-AT-PUB-ASV-2025 BP24-116. H.C.C is also supported by the Severo Ochoa fellowship NAC-AT-PUB-ASV-2025 BP24-35.

\appendix
\section{The 2-point and 1-point functions for a free scalar}\label{sec::2-pointfunction}

The Green's function $G(\vec{x}-\vec{x}')=\langle \phi^{\dagger}(\vec{x})\phi(\vec{x}')\rangle$ in the presence of the defect follows from the equation of motion for $\phi$ --including the background gauge field as described in the main text-- supplemented with the appropriate delta function sources. The equation reads

\begin{equation}
\vec{\partial}^2G+\frac{1}{r}\partial_r\left(r\partial_rG\right)-\frac{1}{r^2}\,(\partial_{\theta}+i e\beta)^2G-m^2G=\frac{1}{2 \pi} \delta(r-r')\delta(\theta-\theta')\delta^2(\vec{\sigma}-\vec{\sigma}')\,.
\end{equation}
Expanding $G$ in Fourier modes

\begin{equation}
G=\int \frac{d^{d-2}\vec{p}}{(2\pi)^{d-2}} \sum_{n\in \mathbb{Z}}\,\frac{1}{2\pi}\,e^{-in(\theta-\theta')}\,e^{i\vec{p}\cdot(\vec{\sigma}-\vec{\sigma}')}\,G_{n,\vec{p}}(r)\,,
\end{equation}
we have equation for the radial modes 

\begin{equation} \label{eq::GreenEqMode}
\partial_r\left(r\partial_rG_{n,\vec{p}}\right)- \left(r \,\omega^2+\frac{(n-e\beta)^2}{r}\right)\,G_{n,\vec{p}}=\,\delta(r-r')\,,\qquad \omega=\sqrt{\vec{p}^2+m^2}\,.
\end{equation}
 Away from $r= r'$ this equation is solved by

\begin{equation}
    G_{n,\vec{p}}(r)= \begin{cases}
        A \,K_{|n-e\beta|}(\omega r)+B \,I_{|n-e\beta|}(\omega r)\quad r<r'\\
        C \,K_{|n-e\beta|}(\omega r)+D \,I_{|n-e\beta|}(\omega r)\quad r>r'\\
    \end{cases} 
\end{equation}
Notice that $ K_{|n-\beta|}(\omega r)$ diverges for $r \to 0$ and $ I_{|n-\beta|}(\omega r)$  diverges for  $r \to  \infty$. We demand that the Green's function is finite in both regions, which sets $D=A=0$. Additionally, imposing that the solution is continuous at t $r=r'$, we find

\begin{equation} \label{eq::contcondG}
B\,K_{|n-e\beta|}(\omega r')=C\,I_{|n-e\beta|}(\omega r')\,.
\end{equation}

Finally, we need to impose the discontinuity produced by the delta function. In order to do so we integrate both sides of \eqref{eq::GreenEqMode} in $r\in (r'-\epsilon,r'+\epsilon)$ for $\epsilon \to 0$,

\begin{equation}
\begin{aligned}
    \lim_{\epsilon\to 0}&\int_{r'-\epsilon}^{r'+\epsilon} dr \left(\, \partial_r\left(r\partial_r G_{n,\vec{p}}\right)- \left(r \,\omega^2+\frac{(n-e\beta)^2}{r}\right)G_{n,\vec{p}}\right)=\lim_{\epsilon\to 0}\left.r\partial_r G_{n,\vec{p}}\right|^{r'+\epsilon}_{r'-\epsilon}\,,\\
     \lim_{\epsilon\to 0}&\int_{r'-\epsilon}^{r'+\epsilon} dr \delta(r-r')=1\,,
\end{aligned}
\end{equation}
where in the first equation the term without the derivative vanishes because we have imposed that $ G_{n,\vec{p}}$ is continuous at $r=r'$. Matching both sides 

\begin{equation}\label{eq::deltacondG}
   r'  \omega C\,I'_{|n-e\beta|}(\omega r') -r'  \omega B K'_{|n-e\beta|}(\omega r')=1\,.
\end{equation}
Solving \eqref{eq::contcondG} and \eqref{eq::deltacondG} we find

\begin{align}
&B=\frac{I_{|n-e\beta|}(\omega r')}{\omega\,r'\,W}\,,\qquad C=\frac{K_{|n-e\beta|}(\omega r')}{\omega\,r'\,W}\,,\nonumber \\
& W=I_{|n-e\beta|}(\omega r')\,K'_{|n-e\beta|}(\omega r')-I'_{|n-e\beta|}(\omega r')\,K_{|n-e\beta|}(\omega r')=-1\,,
\end{align}
being $W$ the Wronskian of the modified Bessel functions. Putting everything together, we can then neatly write

\begin{equation}
G_{n,\vec{p}}(r,r')=-I_{|n-e\beta|}(\omega r_<)\,K_{|n-e\beta|}(\omega r_>)\,,\qquad r_<={\rm min}(r,r')\,,\qquad r_>={\rm max}(r,r')\,.
\end{equation}

\subsection{One-point function}\label{sec::one-pointfunction}

The 1-point function follows from the 2-point function upon substracting the obvious contact divergence. It reads

 \begin{equation}
 \begin{aligned}
     \langle \phi^\dagger \phi(r)\rangle&= \lim _{r'\to r}\langle \phi^\dagger(r) \phi(r')\rangle-\langle \phi^\dagger(r) \phi(r')\rangle_{\beta=0}=\\
     &=\frac{1}{2 \pi}\int \frac{d^{d-2}\vec{p}}{(2\pi)^{d-2}}\, \sum_{n\in \mathbb{Z}}I_{|n-e\beta|}(\omega r)\,K_{|n-e\beta|}(\omega r)-I_{|n|}(\omega r)\,K_{|n|}(\omega r)\,.
 \end{aligned}
 \end{equation}
Making use of the following integral identity

 \begin{equation}
\int_0^{\infty} \frac{ds}{2s}\,e^{-s\omega^2-\frac{r^2}{2s}}\,I_{|\nu|}\left(\frac{r^2}{2s}\right)=I_{|\nu|}(\omega r)\,K_{|\nu|}(\omega r)\,,
\end{equation}
we can write

\begin{equation}
\langle \phi^\dagger \phi(r)\rangle=\int \frac{d^{d-2}\vec{p}}{(2\pi)^{d-1}}\,\int_0^{\infty} \frac{ds}{2s}\,e^{-s\omega^2-\frac{r^2}{2s}} \sum_{n\in \mathbb{Z}}I_{|n-e\beta|}\left(\frac{r^2}{2s}\right)-I_{|n|}\left(\frac{r^2}{2s}\right)\,.
\end{equation}

We can now perform the $\vec{p}$ integral to obtain

\begin{equation} \label{eq::1ptfunctionm}
\langle \phi^\dagger \phi(r)\rangle=\frac{1}{(4 \pi)^\frac{d}{2}}\int_0^{\infty} \frac{ds}{s^{\frac{d}{2}}}\,e^{-sm^2-\frac{r^2}{2s}} \sum_{n\in \mathbb{Z}}I_{|n-e\beta|}\left(\frac{r^2}{2s}\right)-I_{|n|}\left(\frac{r^2}{2s}\right)\,.
\end{equation}

\subsubsection{Zero mass limit}

Upon setting $m=0$, the 1-point function becomes

\begin{equation}
\langle \phi^\dagger \phi(r)\rangle=\frac{1}{(4 \pi)^\frac{d}{2}}\int_0^{\infty} \frac{ds}{s^{\frac{d}{2}}}\,e^{-\frac{r^2}{2s}} \sum_{n\in \mathbb{Z}}I_{|n-e\beta|}\left(\frac{r^2}{2s}\right)-I_{|n|}\left(\frac{r^2}{2s}\right)\,.
\end{equation}
We perform the change of coordinates $x=\frac{r^2}{2s}$

\begin{equation}
\langle \phi^\dagger \phi(r)\rangle=\frac{1}{2(2\pi)^{\frac{d}{2}}r^{d-2}}\, \int_0^{\infty} dx\,x^{\frac{d}{2}-2}\,e^{-x} \sum_{n\in \mathbb{Z}}I_{|n-e\beta|}(x)-I_{|n|}(x)\,.
\end{equation}
Using the following integral representation of the Bessel function,

\begin{equation}\label{eq::intrepIbessel}
I_{\nu}(x)=\frac{1}{2\pi}\int_{-\pi}^{\pi}d\phi\,e^{x\cos\phi}\,e^{i\nu\phi}-\frac{\sin \pi\nu}{\pi}\int_0^{\infty}dt\,e^{-x\cosh t-\nu t}\,,
\end{equation}
the sum becomes

\begin{align}
\sum_{n\in \mathbb{Z}}I_{|n-e\beta|}(x)-I_{|n|}(x)&=\frac{1}{2\pi}\int_{-\pi}^{\pi}d\phi\,e^{x\cos\phi} \sum_n e^{i|n-e\beta|\phi}-e^{i|n|\phi}\\ & \nonumber  -\frac{1}{\pi}\int_0^{\infty}dt\,e^{-x\cosh t}\sum_n \sin (\pi|n-e\beta|)\, e^{-|n-e\beta|t}\,.
\end{align}
Notice that the last term of the second sum drops out since $\sin{\pi |n|}=0$. Now

\begin{equation}
\sin (\pi|n-e\beta|)=\begin{cases} n>0:\qquad \sin (\pi|n-e\beta|)=(-1)^{n+1}\,\sin(e\pi |\beta|)\,,\\ n=0:\,\qquad \sin(e\pi |\beta|)\,,\\ n<0:\qquad (-1)^n\,\sin(e\pi |\beta|)\,.
\end{cases}
\end{equation}
Plugging this into the second sum, we find

\begin{align}
\sum_n \sin (\pi|n-e\beta|)\, e^{-|n-e\beta|t}& =\sin(e\pi |\beta|)\left( \sum_{n\leq 0} (-1)^n\,e^{-(-n-e\beta)t}  + \sum_{n>0} (-1)^{n+1}\,e^{-(n-e\beta)t}\right)\\ \nonumber & =\sin(e\pi \beta)\, \frac{e^{e\beta t}+e^{(1-e\beta)t}}{1+e^t} \,.
\end{align} 
Therefore the full sum becomes

\begin{equation}\label{eq::fullsumI}
\sum_{n\in \mathbb{Z}}I_{|n-e\beta|}(x)-I_{|n|}(x)=\frac{1}{2\pi}\int_{-\pi}^{\pi}d\phi\,e^{x\cos\phi} \sum_n \left( e^{i|n-e\beta|\phi}-e^{i|n|\phi}\right)  - \frac{\sin(e\pi \beta)}{\pi}\int_0^{\infty}dt\,e^{-x\cosh t} \frac{e^{e\beta t}+e^{(1-e\beta)t}}{1+e^t} \,.
\end{equation}

Dropping the first contribution (a renormalization prescription, as discussed in the main text) and recovering all factors

\begin{equation}
   \langle \phi^\dagger \phi(r)\rangle= -\frac{\sin(e\pi \beta)}{(2\pi)^{\frac{d}{2}+}r^{d-2}}\, \int_0^{\infty}dt\, \frac{e^{e\beta t}+e^{(1-e\beta)t}}{1+e^t}\int_0^{\infty}\, dx\,x^{\frac{d}{2}-2} \,e^{-x(1+\cosh t)}  \,.
\end{equation}
The $x$ integral is just a Gamma function,

\begin{equation}
    \langle \phi^\dagger \phi(r)\rangle= -\frac{\sin(e\pi \beta)\, \Gamma(\frac{d-2}{2})}{(2\pi)^{\frac{d}{2}+1}r^{d-2}}\, \int_0^{\infty}dt\,\frac{1}{(1+\cosh t)^{\frac{d-2}{2}}} \frac{e^{e\beta t}+e^{(1-e\beta)t}}{1+e^t}  \,.
\end{equation}

In order to compute the $t$ integral, we reescale the range of the integral to (0,1) by the change of coordinates $u^2=e^{-t}$,

\begin{equation}
    \langle \phi^\dagger \phi(r)\rangle= -\frac{\sin(e\pi \beta)\, \Gamma(\frac{d-2}{2})}{2\pi^{\frac{d}{2}+1}r^{d-2}}\, \int_0^{1}du\,\left(u^2+1\right)^{1-d} \left(u^{4 e\beta}+u^2\right) u^{d-2
  e \beta-3}  \,.
\end{equation}
This integral yields

\begin{equation} \label{eq::2F1ident}
\begin{aligned}
     \int_0^{1}du\,\left(u^2+1\right)^{1-d} \left(u^{4 e\beta}+u^2\right) u^{d-2
  e \beta-3}  =&\frac{\, _2F_1\left(d-1,\frac{d}{2}-e\beta;\frac{d}{2}-e\beta+1;-1\right)}{d-2 e\beta}+\\
   \;&+\frac{\, _2F_1\left(d-1,\frac{d}{2}+e\beta-1;\frac{d}{2}+e\beta;-1\right)}{d+2 e\beta-2}\,. 
\end{aligned}  
\end{equation}
This can be simplified (see Appendix \eqref{sec::hypergeo}) to

\begin{equation} \label{eq::2F1togamma}
     \int_0^{1}du\,\left(u^2+1\right)^{1-d} \left(u^{4 e\beta}+u^2\right) u^{d-2
   e\beta-3}  =\frac{\sqrt{\pi } 2^{2-d} \Gamma
   \left(\frac{d}{2}-e\beta\right) \Gamma
   \left(\frac{d}{2}+e\beta-1\right)}{(d-2) \Gamma
   \left(\frac{d}{2}-1\right) \Gamma
   \left(\frac{d-1}{2}\right)}
\end{equation}
 Now substituting into the full expression we find
 
 \begin{equation}
     \langle \phi^\dagger \phi(r)\rangle=-\frac{ \sin
   (e\pi  \beta) \Gamma \left(\frac{d}{2}-e\beta\right) \Gamma
   \left(\frac{d}{2}+e\beta-1\right)}{2^{d-1}  \pi ^{\frac{d+1}{2}}(d-2) \Gamma
   \left(\frac{d-1}{2}\right)} \frac{1}{r^{d-2}}\,,
 \end{equation}
 which is the result of \cite{Bianchi:2021snj} as expected.

\subsubsection{Large mass limit}
 
Let us now consider the $m\to \infty$ limit. Going back to \eqref{eq::1ptfunctionm} the factor of $e^{-m^2 s}$ suppresses the contributions away from $s=0$. Let us consider on the first sum of \eqref{eq::intrepIbessel},

\begin{equation}
I_{\nu}(x)=\frac{1}{2\pi}\int_{-\pi}^{\pi}d\phi\,e^{x\cos\phi}\,e^{i\nu\phi}\,.
\end{equation}

Since $x=\frac{r^2}{2s}$ is large, there are only contributions near $\phi=0$, since when $\cos \phi$ is negative the integral is suppressed. Similarly, the second sum of \eqref{eq::intrepIbessel} is always suppressed by the factor $e^{-x \cosh t}$. We expand around $\phi=0$,

\begin{equation}
I_{\nu}(x)=\frac{1}{2\pi}e^{x}\int_{-\pi}^{+\pi}d\phi\,e^{-x\frac{\phi^2}{2}}\,e^{i\nu\phi}\sim\frac{e^{x-\frac{\nu^2}{2 x}}}{\sqrt{2 \pi x}}+ \mathcal{O}\left(\frac1{x}\right)\,.
\end{equation}
Then, the difference of Bessel functions becomes
\begin{equation}
\sum_{n\in \mathbb{Z}}I_{|n-e\beta|}\left(\frac{r^2}{2s}\right)-I_{|n|}\left(\frac{r^2}{2s}\right)=\frac{e^{x}}{\sqrt{2 \pi x}}\,\sum_{n\in \mathbb{Z}}e^{-\frac{(n-e\beta)^2}{2x}}-e^{-\frac{n^2}{2x}}\,.
\end{equation}
We substitute this into the full 1-point function

\begin{equation}
\langle \phi^\dagger \phi(r)\rangle=\frac{1}{\sqrt{2 \pi}(4 \pi)^{\frac{d}{2}}r}\, \int_0^{\infty} \frac{ds}{s^{\frac{d-1}{2}}}\,e^{-sm^2} \,\sum_{n\in \mathbb{Z}}e^{-\frac{(n-e\beta)^2}{2x}}-e^{-\frac{n^2}{2x}}\,.
\end{equation}

We now use the following Poisson-Jacobi identities

\begin{equation}
 \sum_{n\in \mathbb{Z}}e^{-\frac{(n-e\beta)^2}{2x}}=\sqrt{2\pi x}\,\sum_{n\in\mathbb{Z}}\,e^{-2x\pi^2 n^2}\,e^{i2\pi n e\beta}\,,\qquad \sum_{n\in \mathbb{Z}}e^{-\frac{n^2}{2x}}=\sqrt{2\pi x}\,\sum_{n\in\mathbb{Z}}\,e^{-2x\pi^2 n^2}\,.
\end{equation}
Then

\begin{equation}
\langle \phi^\dagger \phi(r)\rangle=\frac{1}{(4 \pi)^{\frac{d}{2}}}\,\sum_{n\in\mathbb{Z}}\, \int_0^{\infty} \frac{ds}{s^{\frac{d}{2}}}\,e^{-sm^2-\frac{r^2}{s}\pi^2 n^2}\,(e^{i2\pi ne \beta}-1)\,.
\end{equation}

Let us perform the change of coordinates $z=m s$,

\begin{equation}
\langle \phi^\dagger \phi(r)\rangle=\frac{ m^{\frac{d}{2}-1}}{(4 \pi)^{\frac{d}{2}}}\sum_{n\in\mathbb{Z}}\, \int_0^{\infty} \frac{dz}{z^{\frac{d}{2}}}\,e^{-m\,(z+\frac{r^2}{z}\pi^2 n^2)}\,(e^{i2\pi n e\beta}-1)\,.
\end{equation}

We can now compute the integral in $z$ via saddle-point approximation

\begin{equation}
\langle \phi^\dagger \phi(r)\rangle=\frac{ m^{\frac{d}{2}-1}}{(4 \pi)^{\frac{d}{2}}}\sum_{n\in\mathbb{Z}}\, \int_0^{\infty} \frac{dz}{z^{\frac{d}{2}}}\,e^{-mS}\,(e^{i2\pi ne \beta}-1),\quad  S=(z+\frac{r^2}{z}\pi^2 n^2).
\end{equation}
The critical  action $S_0$ and its Hessian fluctuations $H$ are

\begin{equation}
    \left.\frac{\partial S}{\partial z}\right|_{z_*}=0\implies z_*= n \pi r,\quad S_0=S(z_*)=2 \pi n r,\quad \det H=\left.\frac{\partial^2 S}{\partial z^2}\right|_{z_*}=\frac{2}{n \pi r}\,.
\end{equation}
All in all,

\begin{equation}
\begin{aligned}
    \langle \phi^\dagger \phi(r)\rangle&=\frac{ m^{\frac{d}{2}-1}}{(4 \pi)^{\frac{d}{2}}}\sum_{n\in\mathbb{Z}}\,\frac{\,e^{-mS_0}}{z_*^{\frac{d}{2}} (\det H)^\frac12}\,(e^{i2\pi n e\beta}-1)=\\=& 
  \sqrt{\frac{\pi}{2}} \frac{m^{\frac{d}{2}-1}}{(2 \pi)^d \,r^{\frac{d-1}{2}}}\sum_{n\in\mathbb{Z}}|n|^{\frac{1-d}{2}}
   e^{-2 \pi  m |n| r}(e^{i2\pi ne \beta}-1)\,.
\end{aligned}
\end{equation}
Because of the factor $e^{-2 \pi  m |n| r}$, in the large mass limit the main contribution comes from $n=\pm 1$, so we find

\begin{equation}
    \langle \phi^\dagger \phi(r)\rangle=  \sqrt{\frac{\pi}{2}} \frac{m^{\frac{d}{2}-1}}{(2 \pi)^d } \frac{e^{-2\pi m r}\,\sin^2(e\pi \beta)}{r^{\frac{d-1}{2}}\,}\,.
\end{equation}

\section{Hypergeometric identities}\label{sec::hypergeo}

We start by rewriting the right side of \eqref{eq::2F1ident} as

\begin{equation}
\begin{aligned}
     \frac{\, _2F_1\left(d-1,\frac{d}{2}-\beta;\frac{d}{2}-\beta+1;-1\right)}{d-2 \beta}+\frac{\, _2F_1\left(d-1,\frac{d}{2}+\beta-1;\frac{d}{2}+\beta;-1\right)}{d+2 \beta-2}&= \\
     =\frac{\, _2F_1\left(a,b;b+1;-1\right)}{2b}+\frac{\, _2F_1\left(a,a-b;b-a+1;-1\right)}{2(a-b)}\,,
\end{aligned}
\end{equation}
with $a=d-1,\,b=\frac{d}{2}-\beta$. Now using the following property

\begin{equation}
    _2F_1\left(c,d;d+1;-1\right)= d \,B_{\frac12}(d,c-d)\,,
\end{equation}
where $B_x(c,d)$ is the incomplete beta function defined as

\begin{equation}
    B_x(c,d)=\int_0^x du \;u^{c-1}(1-u)^{d-1},\quad 0\leq x \leq 1\,.
\end{equation}
We can write the previous expression as
\begin{equation}
    \frac{\, _2F_1\left(a,b;b+1;-1\right)}{2b}+\frac{\, _2F_1\left(a,a-b;b-a+1;-1\right)}{2(a-b)}=\frac{1}{2}\left(B_{\frac12}(b,a-b)+B_{\frac12}(a-b,b)\right)\,.
\end{equation}

We will now apply the following property

\begin{equation}
    B_x(p,q)+B_{1-x}(q,p)=B(p,q)\,,
\end{equation}
to finally recover
\begin{equation}
    \frac{1}{2}\left(B_{\frac12}(b,a-b)+B_{\frac12}(a-b,b)\right)=\frac12 B(b,a-b)=\frac{\Gamma(\frac{d}{2}-\beta) \Gamma(\frac{d}{2}+\beta-1)}{2 \Gamma(d-1)}\,.
\end{equation}
After some standard Gamma function transformations one recovers the expression in \eqref{eq::2F1togamma}.

\section{2-point functions in holography}\label{sec::2pointreview}

Let us review the holographic computation of 2-point functions for scalar operators on $\mathbb{R}^{d}$. These are dual to scalar fields of mass $m$ in Poincare $AdS_{d+1}$ in Poincare coordinates

\begin{equation}
S=\frac{1}{2}\int \sqrt{g}\Big(\partial\phi^2+m^2\phi^2\Big)\,,\qquad ds^2=\frac{R^2}{z^2}(dz^2+dx_{1,d-1}^2)
\end{equation}
The equation of motion is

\begin{equation}
z^{d+1}\partial_z(z^{1-d}\partial_z\psi)-z^2\,\vec{p}^2\psi-m^2R^2\psi=0
\end{equation}
Upon Fourier transform, the equation of motion becomes

\begin{equation}
\phi=\int \frac{d\vec{p}}{(2\pi)^d}\,\psi_{\vec{p}}(z)\,e^{i\vec{p}\cdot\vec{x}}\quad \leadsto\quad  z^{d+1}\partial_z(z^{1-d}\partial_z\psi)-z^2\,\vec{p}^2\psi-m^2R^2\psi=0\,.
\end{equation}
This can be easily solved

\begin{equation}
\psi=C_1(\vec{p})\,z^{\frac{d}{2}}\,I_{\nu}(|\vec{p}|z)+C_2(\vec{p})\,z^{\frac{d}{2}}\,K_{\nu}(|\vec{p}|z)\,,\qquad \nu=\sqrt{\frac{d^2}{4}+m^2R^2}\,,\qquad \Delta=\frac{d}{2}+\nu\,.
\end{equation}
Since in the interior $I_{\nu}$ blows up, regularity forces that $C_1(\vec{p})=0$. Then, the solution is

\begin{equation}
\label{sol}
\phi=z^{\frac{d}{2}}\,\int \frac{d\vec{p}}{(2\pi)^d}\,C_2(\vec{p})\,e^{i\vec{p}\cdot\vec{x}}\,K_{\nu}(|\vec{p}|z)\,.
\end{equation}
Close to the boundary, this is

\begin{equation}
\phi=z^{\frac{d}{2}+\nu}\,A(\vec{x})+z^{\frac{d}{2}-\nu}\,B(\vec{x})\,,
\end{equation}
with

\begin{equation}
A(\vec{x})=\int \frac{d\vec{p}}{(2\pi)^d}\,|\vec{p}|^{\nu}\,\frac{\Gamma(-\nu)\,C_2(\vec{p})}{2^{\nu+1}}\,e^{i\vec{p}\cdot\vec{x}}\,,\qquad B(\vec{x})=\int \frac{d\vec{p}}{(2\pi)^d}\,\frac{2^{\nu-1}\Gamma(\nu)\,C_2(\vec{p})}{|\vec{p}|^{\nu}}\,e^{i\vec{p}\cdot\vec{x}}\,.
\end{equation}
Upon Fourier-transforming $B(\vec{x})$ we can identify $C_2(\vec{p})$

\begin{equation}
B(\vec{x})=\int \frac{d\vec{p}}{(2\pi)^d} B(\vec{p})\,e^{i\vec{p}\cdot\vec{x}}\,,\qquad B(\vec{p})=\frac{2^{\nu-1}\Gamma(\nu)\,C_2(\vec{p})}{|\vec{p}|^{\nu}}\,.
\end{equation}
Moreover

\begin{equation}
\label{AB}
A(\vec{x})=\int \frac{d\vec{p}}{(2\pi)^d}\,\frac{\Gamma(-\nu)}{2^{2\nu}\,\Gamma(\nu)}\,|\vec{p}|^{2\nu}\,B(\vec{p})\,e^{i\vec{p}\cdot\vec{x}}\,.
\end{equation}

Turning now to the on-shell action, we find

\begin{equation}
S_{\rm os}=\int d\vec{x} z^{1-d}\,\frac{1}{2}\phi\partial_z\phi\Big|_{z=\epsilon}=\int d\vec{x}\,\frac{d}{2}\,A\,B+\frac{d-2\nu}{4}\,B^2\,\epsilon^{-2\nu}\,,
\end{equation}
where $\epsilon$ is a boundary regulator. To renormalize the divergences, we add the counterterm

\begin{equation}
S_{\rm ct}=-\frac{d-2\nu}{4}\int d\vec{x} \sqrt{g_{\partial}}\,\phi^2\Big|_{z=\epsilon}=\int d\vec{x}\,\left(-\frac{d-2\nu}{2}\,A\,B-\frac{d-2\nu}{4}B^2\,\epsilon^{-2\nu}\right)\,.
\end{equation}
Hence the renormalized action $S_{\rm ren}=S_{\rm os}+S_{\rm ct}$ is finite and reads

\begin{equation}
S_{\rm ren}=\int d\vec{x}\,\nu\,A(\vec{x})\,B(\vec{x})=\frac{\nu\,\Gamma(-\nu)}{4^{\nu}\,\Gamma(\nu)}\,\int \frac{d\vec{p}}{(2\pi)^d}  B(-\vec{p})\,B(\vec{p})\, |\vec{p}|^{2\nu} \,.
\end{equation}
We can now use a standard Fourier-transform identity

\begin{equation}
\frac{1}{|\vec{p}|^{-2\nu}}=\frac{(4\pi)^{\frac{d}{2}}\,\Gamma(\frac{d}{2}-\nu)}{(2\pi)^d\,4^{\nu}\,\Gamma(\nu)}\,\int d\vec{x}\,\frac{e^{i\vec{p}\cdot\vec{x}}}{|\vec{x}|^{d+2\nu}}\,.
\end{equation}
Then

\begin{equation}
S_{\rm ren}=\mathcal{N} \int d\vec{x} \int d\vec{y} \,B(\vec{x})\,\frac{1}{|\vec{x}-\vec{y}|^{d+2\nu}}\,B(\vec{y}) \,,
\end{equation}
with $\mathcal{N}$ a numerical factor. Thus, we see that the generating functional for correlation functions of the scalar operator is $e^{-S_{\rm ren}}$, so that functional derivatives with respect to $B(\vec{x})$ give the corresponding correlators. Note that a shortcut to obtain those stems from \eqref{AB}, which in momentum space for $B$ it reads

\begin{equation}
\label{ABmomentum}
A(\vec{p})=\frac{\Gamma(-\nu)}{2^{\nu}\Gamma(\nu)}\,|\vec{p}|^{2\nu}\,B(\vec{p})\,.
\end{equation}
Hence, normalization aside, we can read-off the 2-point function from the ratio of subleading-to-leading modes.

\subsection{Another approach}

Let us suppose that we want to solve the equation of motion with the Dirichlet boundary condition that $\phi$ approaches $\phi_0(\vec{x})$ at the boundary. On general grounds, we could write

\begin{equation}
\label{phiintermsofK}
\phi=\int d\vec{y}\, \phi_0(\vec{y})\,K(\vec{x},z;\vec{y})\,,
\end{equation}
where $K(\vec{x},z;\vec{y})$ is the boundary-to-bulk propagator defined as the Green's function of the bulk equation of motion with the source pushed to the boundary. In the vanilla $AdS_5$ case, it turns out that

\begin{equation}
K(\vec{x},z;\vec{y})=\left(\frac{z}{(\vec{x}-\vec{y})^2+z^2}\right)^{\Delta}\,.
\end{equation}

To connect with the previous description, let us use the Fourier transform identity

\begin{equation}
\label{FourierTransform}
\left(\frac{z}{\vec{x}^2+z^2}\right)^{\Delta}=z^{\frac{d}{2}}\,\int d\vec{p}\,\frac{|\vec{p}|^{\nu}\,K_{\nu}(|\vec{p}|z)}{2^{\Delta-1}\,(2\pi)^{\frac{d}{2}}\,\Gamma(\Delta)}\,e^{i\vec{p}\cdot\vec{x}}\,.
\end{equation}
Then we find exactly the form of the solution in \eqref{sol}

\begin{equation}
\phi=z^{\frac{d}{2}}\,\int \frac{d\vec{p}}{(2\pi)^d}\, \frac{B(\vec{p})}{2^{\nu-1}\,\Gamma(\nu)}\,|\vec{p}|^{\nu}\,e^{i\vec{p}\cdot\vec{x}}\, K_{\nu}(|\vec{p}|z)\,.
\end{equation}
where

\begin{equation}
B(\vec{p})=\frac{(2\pi)^{\frac{d}{2}}\,2^{\nu-1}\Gamma(\nu)}{2^{\Delta-1}\,\Gamma(\Delta)}\,\int d\vec{y}\,\phi_0(\vec{y})\,e^{-i\vec{p}\vec{y}}\,.
\end{equation}

\subsubsection{Boundary-to-bulk propagator}

From this point of view the computation boils down to computing the boundary-to-bulk propagator $K$. We stress that, in order to compute 2-point functions, only its behavior close to the boundary is relevant. Assuming asymptotically $AdS$ spaces --as it is the case at hand--, it is enough then to consider the the bulk equation of motion in $AdS_{d+1}$. Suppose adding a $\delta$ source to the equation of motion. The Green's function equation is\footnote{Note the RHS normalization of the $\delta(z-z')$.}

\begin{equation}
z^{d+1}\partial_z(z^{1-d}\partial_zG)+z^2\,\vec{\partial}^2G-m^2R^2G=z^{d+1}\,\delta(\vec{x}-\vec{y})\delta(z-z')\,.
\end{equation}
Writting

\begin{equation}
G=\int \frac{d\vec{p}}{(2\pi)^d}\,\psi\,e^{i\vec{p}\cdot\vec{x}}\,,
\end{equation}
the equation to solve is

\begin{equation}
z^{d+1}\partial_z(z^{1-d}\partial_z\psi)-z^2\,\vec{p}^2\psi-m^2R^2\psi=z^{d+1}\,\delta(z-z')\,.
\end{equation}
Away from coincident points, the solution is the familar Bessel function $z^{\frac{d}{2}}\,I_{\nu}(|\vec{p}|z)$ or $z^{\frac{d}{2}}\,K_{\nu}(|\vec{p}|z)$. For $z\rightarrow \infty$ the regular solution is that with $K_{\nu}$. Thus we write

\begin{equation}
\psi=\begin{cases} z<z':\qquad A\,z^{\frac{d}{2}}\,I_{\nu}(|\vec{p}|z),\\ z>z' \qquad B\,z^{\frac{d}{2}}\,K_{\nu}(|\vec{p}|z)\,.\end{cases}
\end{equation}
The continuity of the function requires

\begin{equation}
A\,I_{\nu}(|\vec{p}|z')=B\,K_{\nu}(|\vec{p}|z')\,.
\end{equation}
In turn, integrating the function in a small neigbourhood of $z'$ gives

\begin{equation}
B\partial_z(z^{\frac{d}{2}}K_{\nu}(|\vec{p}|z))-A\,\partial_z(z^{\frac{d}{2}}\,I_{\nu}(|\vec{p}|z))=z^{d-1}\,,
\end{equation}
all evaluated at $z=z'$. This fixes

\begin{equation}
A=-z'^{-\frac{d}{2}}\,K_{\nu}(|\vec{p}|z')\,,\qquad B=-z'^{-\frac{d}{2}}\,I_{\nu}(|\vec{p}|z')\,.
\end{equation}
Thus

\begin{equation}
\psi=\begin{cases} z<z':\qquad -K_{\nu}(|\vec{p}|z')\,(z\,z')^{\frac{d}{2}}\,I_{\nu}(|\vec{p}|z),\\ z>z' \qquad -I_{\nu}(|\vec{p}|z')\,(z\,z')^{\frac{d}{2}}\,K_{\nu}(|\vec{p}|z)\,.\end{cases}
\end{equation}

Let us now send $z'$ to the boundary. For small argument

\begin{equation}
I_{\nu}(|\vec{p}|z')=\frac{1}{2^{\nu}\,\Gamma(\nu+1)}|\vec{p}|^{\nu}\,z'^{\nu}\,.
\end{equation}
Thus we are always in the $z>z'$ regime and

\begin{equation}
\psi= -\frac{z'^{\Delta}}{2^{\nu}\,\Gamma(\nu+1)}\,z^{\frac{d}{2}}\,|\vec{p}|^{\nu}\,K_{\nu}(|\vec{p}|z)\,.
\end{equation}
Thus, the Green's function goes over to (let us denote it as $G_{\partial b}$ to remember that we have sent $z'$ to the boundary and kept $z$ in the bulk)

\begin{equation}
G_{\partial b}=-\frac{z'^{\Delta}}{2^{\nu}\,\Gamma(\nu+1)}\,z^{\frac{d}{2}}\,\int \frac{d\vec{p}}{(2\pi)^d}\,|\vec{p}|^{\nu}\,K_{\nu}(|\vec{p}|z)\,e^{i\vec{p}\cdot\vec{x}}\,.
\end{equation}
Using the Fourier transform identity \eqref{FourierTransform}

\begin{equation}
G_{\partial b}=-\frac{z'^{\Delta}}{2^{\nu}\,\Gamma(\nu+1)}\,\frac{2^{\Delta-1}\Gamma(\Delta)}{(2\pi)^{\frac{d}{2}}}\,K(\vec{x},z;\vec{y})\,.
\end{equation}
So we can write

\begin{equation}
\label{K}
K(\vec{x},z;\vec{y})=\mathcal{N}\,\lim_{z'\rightarrow 0} z'^{-\Delta}\,G(z,\vec{x};z',\vec{y})\,.
\end{equation}
where $\mathcal{N}$ is a numerical factor.

\subsubsection{The renormalized action in terms of Green's functions}

Using \eqref{phiintermsofK} and \eqref{K}, we can write the solution in terms of the Green's function for the bulk equation of motion as

\begin{equation}
\phi(\vec{x},z)=\mathcal{N}\,\lim_{z'\rightarrow 0}z'^{-\Delta}\, \int d\vec{y}\, \phi_0(\vec{y})\,G(z,\vec{x};z',\vec{y})\,.
\end{equation}
The on-shell action would then be

\begin{equation}
S_{\rm os}=\mathcal{N}^2\,\lim_{z\rightarrow 0}\lim_{z',z''\rightarrow 0}\,z'^{-\Delta}z''^{-\Delta}\,\int d\vec{x}\int d\vec{y}_1\int d\vec{y}_2\,\phi_0(\vec{y}_1)\,\phi_0(\vec{y}_2)\,G(z,\vec{x};z',\vec{y}_1)\,\frac{z^{-d+1}}{2}\,\partial_zG(z,\vec{x};z'',\vec{y}_2)\, .
\end{equation}
The counterterm action is in turm

\begin{equation}
S_{\rm ct}=\mathcal{N}^2\,\lim_{z\rightarrow 0}\lim_{z',z''\rightarrow 0}\,z'^{-\Delta}z''^{-\Delta}\, \int d\vec{x} \int d\vec{y}_1\int d\vec{y}_2\, \phi_0(\vec{y}_1)\,\phi_0(\vec{y}_2)\,G(z,\vec{x};z',\vec{y}_1)\,\left( -\frac{d-2\nu}{4}z^{-d}\right) G(z,\vec{x};z'',\vec{y}_2)
\end{equation}
Thus the renormalized action is

\begin{equation}
\label{eq::osactiontouse}
S_{\rm ren}=\frac{1}{2}\,\mathcal{N}^2\,\int d\vec{y}_1\int d\vec{y}_2\, \phi_0(\vec{y}_1)\,\phi_0(\vec{y}_2)\,\mathcal{G}(\vec{y}_1,\vec{y}_2)\,,
\end{equation}
where

\begin{equation}
\mathcal{G}(\vec{y}_1,\vec{y}_2)= \lim_{z\rightarrow 0}\lim_{z',z''\rightarrow 0}\,z'^{-\Delta}z''^{-\Delta}\, \int d\vec{x}\,  G(z,\vec{x};z',\vec{y}_1)\,\left( z^{-d+1}\,\partial_z-\frac{d-2\nu}{2}z^{-d}\right) G(z,\vec{x};z'',\vec{y}_2)\,.
\end{equation}
This shows that the 2-point function is directly $\mathcal{G}(\vec{y}_1,\vec{y}_2)$. Using now Green's identity (see \eqref{Green}), it can be written completely in terms of the Green's function of the bulk equation of motion

\begin{equation}
\label{eq:limitG}
\mathcal{G}(\vec{x}',\vec{x})=\lim_{z,z'\rightarrow 0}\,z^{-\Delta}\,z'^{-\Delta}\,G(z',\vec{x}';z,\vec{x})\,.
\end{equation}

\subsection{Proof of Green's identity}\label{Green}

We want to prove

\begin{equation}
\mathcal{G}(\vec{y}_1,\vec{y}_2)= \lim_{z,z_1,z_2\rightarrow 0}\,z_1^{-\Delta}z_2^{-\Delta}\, \int d\vec{x}\,  G(z,\vec{x};z_1,\vec{y}_1)\,\mathcal{D} G(z,\vec{x};z_2,\vec{y}_2)\,,\qquad  \mathcal{D}= z^{-d+1}\,\partial_z-\frac{d-2\nu}{2}z^{-d}\,.
\end{equation}

Here $G$ is a Green's function satisfying

\begin{equation}
\mathcal{L}G(z,\vec{x};z',\vec{x}')=z^{d+1}\,\delta(\vec{x}-\vec{x}')\delta(z-z')\,,
\end{equation}
where we introduced the operator $\mathcal{L}$ 

\begin{equation}
\mathcal{L}f=z^{d+1}\partial_z(z^{1-d}\partial_zf)+z^2\,\vec{\partial}^2f-m^2R^2f\,.
\end{equation}

We now do the usual thing

\begin{equation}
z^{-d-1}\left(u\mathcal{L}-v\mathcal{L}v\right)=\partial_z\left( z^{1-d}(u\partial_z v-v\partial_z u)\right)+\vec{\partial}\left(z^{1-d}\,(u \vec{\partial}v-v\vec{\partial}u)\right)\,.
\end{equation}
We can write this as

\begin{equation}
z^{-d-1}\left(u\mathcal{L}-v\mathcal{L}v\right)=\partial_I\left( z^{1-d}(u\partial_I v-v\partial_I u)\right)\,.
\end{equation}
Integrating over some space $\mathcal{M}$

\begin{equation}
\int_{\mathcal{M}}z^{-d-1}\left(u\mathcal{L}-v\mathcal{L}v\right)=\int_{\partial\mathcal{M}} \, z^{1-d}(u\partial_n v-v\partial_n u)\,.
\end{equation}
Let us now use this with $u=G(X,Y_1)$, $v=G(X,Y_2)$, where we have eased notation defining $X=(z,\vec{x})$, $Y_i=(z_i,\vec{y}_i)$, we have

\begin{equation}
z^{-d-1}\,\mathcal{L}G(X,Y_i)=\delta(X-Y_i)\,,
\end{equation}
so

\begin{align}
&\int_{\mathcal{M}}\, G(X,Y_1)\delta(X-Y_2)-G(X,Y_2) \delta(X-Y_1) \nonumber \\  &=\int_{\partial\mathcal{M}}\, G(X,Y_1) (z^{1-d}\partial_n) G(X,Y_2)-G(X,Y_2)(z^{1-d}\partial_n) G(X,Y_1)\,.
\end{align}
We now need to specify $\mathcal{M}$. If we choose it to be the whole space $\widehat{\mathcal{M}}=\mathbb{R}^d\times \mathbb{R}^+$, so that both $Y_1$ and $Y_2$ are contained, the LHS gives 0.\footnote{Recall that $G(X,Y)=G(Y,X)$.} Thus, we will choose $\mathcal{M}=\widehat{\mathcal{M}}-B_{\rho}(Y_1)$, where $B_{\rho}(Y_1)$ is a radius $\rho$ (which eventually we will want to send to zero) ball around $Y_1$. Then 

\begin{align}
\nonumber G(Y_1,Y_2)&=\lim_{z\rightarrow 0} \int_{\mathbb{R}^d}\, G(X,Y_1) (z^{1-d}\partial_z) G(X,Y_2)-G(X,Y_2)(z^{1-d}\partial_z) G(X,Y_1)\\ & +\int_{\partial B_{\rho}(Y_1)}\, G(X,Y_1) (z^{1-d}\partial_n) G(X,Y_2)-G(X,Y_2)(z^{1-d}\partial_n) G(X,Y_1)\,.
\end{align}
where $n$ stands for the normal vector to the ball surface.

Now, for the ball integration, we can schematically write $X=Y_1+\delta$, so that the integration is over $\delta$ constrained to live on the boundary of the ball. But then $G(X,Y_2)\sim G(Y_1,Y_2)+\cdots$. The $z$-derivative of this is some regular function and its integral over the ball will be proportional to the area of the boundary of the ball. Thus, in the limit $\rho\rightarrow 0$ the first term gives a vanishing contribution. Thus we can keep

\begin{align}
\nonumber G(Y_1,Y_2)&=\lim_{z\rightarrow 0} \int_{\mathbb{R}^d}\, G(X,Y_1) (z^{1-d}\partial_z) G(X,Y_2)-G(X,Y_2)(z^{1-d}\partial_z) G(X,Y_1)\\ & -G(Y_1,Y_2)\,\int_{\partial B_{\rho}(Y_1)}\, (z^{1-d}\partial_n) G(X,Y_1)\,.
\end{align}
To evaluate the last term let us consider the equation for the Green's function

\begin{equation}
z^{d+1}\partial_z(z^{1-d}\partial_zG(X,Y_1))+z^2\,\vec{\partial}^2G(X,Y_1)-m^2R^2G(X,Y_1)=z^{d+1}\,\delta(X-Y_1)\,.
\end{equation}
Integrating this equation on $B_{\rho}(Y_1)$

\begin{equation}
\int_{B_{\rho}(Y_1)}\partial_I(z^{1-d}\partial_IG(X,Y_1))-z^{-d-1}m^2R^2G(X,Y_1)=1\,.
\end{equation}
Since the ball is very small, $X\sim Y_1$, so in the mass-term we can write approximately $G(X,Y_1)\sim |X-Y_1|^{1-d}= \rho^{1-d}$. On the other hand, the ball-volume element goes with $d\rho\,\rho^d$. Thus, all together, the mass-term will give $\int d\rho \rho\rightarrow 0$. Then we are left with

\begin{equation}
\int_{\partial B_{\rho}(Y_1)}\,z^{1-d}\partial_nG(X,Y_1))=1\,.
\end{equation}
Thus, coming back to our computation

\begin{equation}
G(Y_1,Y_2)=\frac{1}{2}\,\lim_{z\rightarrow 0} \int_{\mathbb{R}^d}\, G(X,Y_1) (z^{1-d}\partial_z) G(X,Y_2)-G(X,Y_2)(z^{1-d}\partial_z) G(X,Y_1)\,.
\end{equation}
Let us now do

\begin{equation}
G(Y_1,Y_2)=\lim_{z\rightarrow 0} \int_{\mathbb{R}^d}\, G(X,Y_1) (z^{1-d}\partial_z) G(X,Y_2)-\frac{1}{2}\,\lim_{z\rightarrow 0} \int_{\mathbb{R}^d}\,z^{1-d}\,\partial_z(G(X,Y_2) G(X,Y_1))\,.
\end{equation}
Now, as we have argued, near the boundary $G\sim z^{\Delta}\,K$. As a consequence, the second term simply vanishes. Thus we finally have

\begin{equation}
G(Y_1,Y_2)=\lim_{z\rightarrow 0} \int_{\mathbb{R}^d}\, G(X,Y_1) (z^{1-d}\partial_z) G(X,Y_2)\,.
\end{equation}

Here we have been a bit cavalier with the treatment of the $z\rightarrow 0$ limit. Strictly speaking, we should have regulated $z=\epsilon$ and added the counterterms. Then

\begin{equation}
G(Y_1,Y_2)=\lim_{z\rightarrow 0} \int_{\mathbb{R}^d}\, G(X,Y_1) \mathcal{D}G(X,Y_2)\,,
\end{equation}
as desired.

\section{Review of the heat kernel method} \label{sec::heatkern}

The heat kernel method can be used to solve equations of the type

\begin{equation}
\Box_x \phi(x)=J(x)\,
\end{equation}
for some differential operator $\Box_x$ depending on some coordinates $x$. Formally, the solution to this equation is $\phi(x)=\Box_x^{-1}J(x)$. We may express the inverse of the differential operator as
\begin{equation}
\label{opid}
\Box_x^{-1}=\int_0^{\infty}ds\,e^{-s\Box_x}\,.
\end{equation}
Indeed, this comes from
\begin{equation}
\frac{d}{ds}(e^{-s\Box_x})=-\Box_x\,e^{-s\Box_x}\qquad\leadsto \qquad \int_0^{\infty}ds \frac{d}{ds}(e^{-s\Box_x})=-1=-\Box_x\,\int_0^{\infty}ds\,e^{-s\Box_x}\,,
\end{equation}
and so the identity \eqref{opid} follows.

The action of the operator can be considered in position space

\begin{equation}
\label{heatkernel}
e^{-s\Box_x}f(x)=\int dy\,K(s,x,y)\,f(y)\,,
\end{equation}
where we dub $K(s,x,y)$ as the heat kernel. Then we can write the solution to our original equation

\begin{equation}
\phi(x)=\Box_x^{-1}J(x)=\int dy\,\int_0^{\infty}ds\, K(s,x,y)\,f(y)\,.
\end{equation}
In particular we see that the Green's function is

\begin{equation}
G(x,y)=\int_0^{\infty}ds\, K(s,x,y)\,.
\end{equation}

Finally, take $U=e^{-s\Box_x}$. Then $U$ satisfies

\begin{equation}
(\partial_s+\Box_x)U=0\,.
\end{equation}
Now let us act with this equation on a function $f(x)$. Denoting $u(s,x)=Uf(x)$, we find

\begin{equation}
(\partial_s+\Box_x)u(t,x)=0\,.
\end{equation}
We impose that $u(0,x)=f(x)$. Using now  \eqref{heatkernel}, the equation is

\begin{equation}
(\partial_s+\Box_x)\int dy\,K(s,x,y)f(y)=0\,,
\end{equation}
From here it follows that

\begin{equation}
(\partial_s+\Box_x)\,K(s,x,y)=0\,,\qquad K(0,x,y)=\delta(x-y)\,.
\end{equation}
The last condition is to ensure the boundary condition $u(0,x)=f(x)$.
\subsection{Coincident limit at $s \to 0$}
Let us now consider taking $y \to x$. Let us consider the heat kernel of the 1d harmonic oscillator, which

\begin{equation}
K(x,x',s)=\sqrt{\frac{\omega}{2\pi \sinh(2\omega s)}}\,e^{-\frac{\omega}{2\sinh(2\omega s)}\left((x^2+x'^2)\cosh (2\omega s)-2xx' \right)}
\end{equation}
Taking $s\rightarrow 0$

\begin{equation}
K=\frac{1}{2\sqrt{\pi}\sqrt{s}}\,e^{-\frac{(x-x')^2}{4s}}\,e^{-\frac{x^2+x'^2+xx'}{3}\omega^2\,s}\,.
\end{equation}
So if $x=x'$

\begin{equation}
K=\frac{1}{2\sqrt{\pi}\sqrt{s}}\,e^{-s\omega^2x^2}\,.
\end{equation}
The heat kernel reduces to the exponential of the potential $V= \omega^2 x^2$ and a prefactor $\sim s^{-\frac{1}{2}}$ which at the $s\to 0$ limit gives the expected delta function behavior.

Let us now consider a heat equation closer to our WKB problem,
\begin{equation}
(\partial_s+F(y)\partial_y(G(y)\partial_y)+V(y))K=0\,.
\end{equation}
We now do the coordinate change $K={Q}/{\sqrt{G}}$.
\begin{equation}
(\partial_s+F\,G\,\partial_y^2+U)Q=0\,,\qquad U=V-\frac{F}{2}\left(G''+\frac{G'^2}{2G}\right)\,.
\end{equation}
Let us define a new coordinate
\begin{equation}
\frac{d\rho}{dy}=\frac{1}{\sqrt{FG}}\,,\quad\partial_y = \frac{1}{\sqrt{FG}}\partial_{\rho}\,,\,\quad \partial_y^2=\frac{1}{FG}\left(\partial_{\rho}^2-\frac{1}{2\sqrt{FG}}\frac{d(FG)}{dy}\partial_{\rho}\right)\,,
\end{equation}
so the equation becomes
\begin{equation}
(\partial_s+\partial_{\rho}^2-\frac{1}{2\sqrt{FG}}\frac{d(FG)}{dy}\partial_{\rho}+U)Q=0\,.
\end{equation}
Define now 

\begin{equation}
Q=e^{B(\rho)}L\,,\qquad \frac{dB}{d\rho}=\frac{1}{4\sqrt{FG}}\frac{d(FG)}{dy}\,.
\end{equation}
Then

\begin{equation}
(\partial_s+\partial_{\rho}^2+W)L=0\,,\qquad W=U+\frac{1}{4}\partial_{\rho}^2\log(FG)-\frac{1}{16}(\partial_{\rho}\log(FG))^2\,.
\end{equation}
Putting it all together, we can write

\begin{equation}
W=V+\frac{1}{4}\partial_{\rho}^2\log(FG)-\frac{1}{2} \partial_{\rho}^2\log G-\frac{1}{2}(\partial_{\rho}\log G)^2-\left(\frac{1}{2}\partial_{\rho}\log G-\frac{1}{4}\partial_{\rho}\log(FG)\right)^2\,.
\end{equation}

Let us then try to solve 

\begin{equation}
(\partial_s+\partial_{\rho}^2+W)L=0\,,
\end{equation}
in the $s\to 0$ limit. In order to do so, let us re-scale $s=\epsilon\tau$ and $\rho=\rho_0+\sqrt{\epsilon}\,r$, so that the equation is
\begin{equation}
(\partial_{\tau}+\partial_r^2+\epsilon\,W)L=0\,.
\end{equation}
We now expand $W=W(\rho_0+\sqrt{\epsilon} r)$ for small $\epsilon$, so at leading order
\begin{equation}
(\partial_{\tau}+\partial_r^2+\epsilon\,W_0)L=0\,,
\end{equation}
where $W_0=W(\rho_0)$. The solution to this is

\begin{equation}
L=\frac{1}{\sqrt{4\pi \tau}}\,e^{\frac{(r-r')^2}{4\tau}}\,e^{-s\epsilon W_0}\,.
\end{equation}
Let us now assume that the functions $F$, $G$ are slowly varying at $y_0$. Then we can neglect all derivatives and at the end, for coincident points

\begin{equation}
K=\frac{1}{\sqrt{4\pi s}}\,e^{-s V(y)}\,,
\end{equation}
where $y$ (the former $\rho_0$) is understood as a fixed value.

It should be stressed that in the adiabatic approximation, since we are missing


\begin{thebibliography}{99}


\bibitem{Rodriguez-Gomez:2026mjj}
D.~Rodriguez-Gomez,
``Holographic Correlators of Giant Gravitons in Monodromy Defects,''
[arXiv:2601.10788 [hep-th]].

\bibitem{Bianchi:2021snj}
L.~Bianchi, A.~Chalabi, V.~Proch{\'a}zka, B.~Robinson and J.~Sisti,
``Monodromy defects in free field theories,''
JHEP \textbf{08} (2021), 013
doi:10.1007/JHEP08(2021)013
[arXiv:2104.01220 [hep-th]].

\bibitem{Arav:2024exg}
I.~Arav, J.~P.~Gauntlett, Y.~Jiao, M.~M.~Roberts and C.~Rosen,
``Superconformal monodromy defects in $ \mathcal{N} $=4 SYM and LS theory,''
JHEP \textbf{08} (2024), 177
doi:10.1007/JHEP08(2024)177
[arXiv:2405.06014 [hep-th]].

\bibitem{Bomans:2024vii}
P.~Bomans and L.~Tranchedone,
``Holographic generalised Gukov-Witten defects,''
JHEP \textbf{03} (2025), 118
doi:10.1007/JHEP03(2025)118
[arXiv:2410.18172 [hep-th]].

\bibitem{Conti:2025qwn}
A.~Conti and R.~Stuardo,
``Monodromy Defects in Maximally Supersymmetric Yang-Mills Theories from Holography,''
[arXiv:2512.10767 [hep-th]].


\bibitem{Conti:2025wyj}
A.~Conti, Y.~Lozano and C.~Rosen,
``Monodromy defects in massive Type IIA,''
JHEP \textbf{04} (2026), 173
doi:10.1007/JHEP04(2026)173
[arXiv:2512.10006 [hep-th]].

\bibitem{Conti:2025wwf}
A.~Conti, Y.~Lozano, F.~Rogdakis and C.~Rosen,
``Defect entanglement entropy for superconformal monodromy defects,''
JHEP \textbf{05} (2026), 036
doi:10.1007/JHEP05(2026)036
[arXiv:2511.22695 [hep-th]].


\bibitem{Copetti:2026ncv}
C.~Copetti,
``When Symmetries Twist: Anomaly Inflow on Monodromy Defects,''
[arXiv:2605.16482 [hep-th]].

\bibitem{Gomis:2025gzb}
J.~Gomis,
``The AdS/$\mathsf{C}$-$\mathsf{P}$-${\mathsf T}$ Correspondence,''
[arXiv:2507.12467 [hep-th]].

\bibitem{Linardopoulos:2026mut}
G.~Linardopoulos and C.~Park,
``Heavy holographic correlators in defect conformal field theories,''
[arXiv:2601.15736 [hep-th]].

\bibitem{Georgiou:2023yak}
G.~Georgiou, G.~Linardopoulos and D.~Zoakos,
``Holographic correlators of semiclassical states in defect CFTs,''
Phys. Rev. D \textbf{108} (2023) no.4, 046016
doi:10.1103/PhysRevD.108.046016
[arXiv:2304.10434 [hep-th]].

\bibitem{Bissi:2011dc}
A.~Bissi, C.~Kristjansen, D.~Young and K.~Zoubos,
``Holographic three-point functions of giant gravitons,''
JHEP \textbf{06} (2011), 085
doi:10.1007/JHEP06(2011)085
[arXiv:1103.4079 [hep-th]].

\bibitem{Holguin:2025dei}
A.~Holguin,
``Semiclassics, branes, and extremality,''
[arXiv:2512.24979 [hep-th]].

\bibitem{Anempodistov:2026dhi}
P.~Anempodistov,
``Holographic two-point functions of heavy operators revisited,''
[arXiv:2603.28880 [hep-th]].

\bibitem{Barkeshli:2025cjs}
M.~Barkeshli, C.~Fechisin, Z.~Komargodski and S.~Zhong,
``Disclinations, Dislocations, and Emanant Flux at Dirac Criticality,''
Phys. Rev. X \textbf{16} (2026) no.1, 011017
doi:10.1103/kfd3-qtk7
[arXiv:2501.13866 [cond-mat.str-el]].

\bibitem{Drukker:2008wr}
N.~Drukker, J.~Gomis and S.~Matsuura,
JHEP \textbf{10} (2008), 048
doi:10.1088/1126-6708/2008/10/048
[arXiv:0805.4199 [hep-th]].

\bibitem{Choi:2024ktc}
C.~Choi, J.~Gomis and R.~Izquierdo Garc{\'\i}a,
``Surface operators and exact holography,''
JHEP \textbf{12} (2024), 195
doi:10.1007/JHEP12(2024)195
[arXiv:2406.08541 [hep-th]].

\bibitem{IzquierdoGarcia:2025jyb}
R.~Izquierdo Garc{\i}a,
``Higher dimensional holography,''
[arXiv:2512.12696 [hep-th]].






\bibitem{Kunduri:2007qy}
H.~K.~Kunduri and J.~Lucietti,
``Near-horizon geometries of supersymmetric AdS(5) black holes,''
JHEP \textbf{12} (2007), 015
doi:10.1088/1126-6708/2007/12/015
[arXiv:0708.3695 [hep-th]].

\bibitem{Ferrero:2021etw}
P.~Ferrero, J.~P.~Gauntlett and J.~Sparks,
``Supersymmetric spindles,''
JHEP \textbf{01} (2022), 102
doi:10.1007/JHEP01(2022)102
[arXiv:2112.01543 [hep-th]].






\end{thebibliography}
\end{document}